\documentclass[twocolumn,reprint]{revtex4}
\usepackage{physics}
\usepackage{enumitem}
\usepackage{graphicx}
\usepackage{color}
\usepackage{hyperref}
\usepackage{amsmath}
\newcommand{\br}{\bm{r}}
\newcommand{\bk}{\bm{k}}
\newcommand{\be}{\begin{equation}}
\newcommand{\ee}{\end{equation}}

\RequirePackage{amsmath,amsfonts,amssymb,bm,braket}
\usepackage[normalem]{ulem}

\usepackage{graphicx}

\begin{document}

\title{Computational methods for 2D materials modelling}
\author{ A. Carvalho$^\dagger$}
\affiliation{Centre for Advanced 2D Materials, National University of Singapore, 6 Science Drive 2, 117546 Singapore}
\email{carvalho@nus.edu.sg}
\author{P. E. Trevisanutto$^\dagger$}
\affiliation{European Centre for Theoretical Studies in Nuclear Physics and Related Areas (ECT*-FBK) and Trento Institute for Fundamental Physics and Applications
(TIFPA-INFN), Via Sommarive, 14, 38123 Povo TN, Trento, Italy}
\author{S. Taioli$^\dagger$} 
\affiliation{European Centre for Theoretical Studies in Nuclear Physics and Related Areas (ECT*-FBK) and Trento Institute for Fundamental Physics and Applications
(TIFPA-INFN), Via Sommarive, 14, 38123 Povo TN, Trento, Italy}
\affiliation{Peter the Great St. Petersburg Polytechnic University, Polytechnicheskaya 29, St. Petersburg 195251, Russia}
\author{A. H. Castro Neto}
\affiliation{Centre for Advanced 2D Materials, National University of Singapore, 6 Science Drive 2, 117546 Singapore}
\affiliation{Materials Science and Engineering, National University of Singapore, 9 Engineering Drive 1, 117575 Singapore }

\begin{abstract}
 Materials with thickness ranging from a few nanometers to a single atomic layer present unprecedented opportunities to investigate new phases of matter constrained to the two-dimensional plane. Particle-particle Coulomb interaction is dramatically affected and shaped by the dimensionality reduction, driving well-established solid state theoretical approaches to their limit of applicability. Methodological developments in theoretical modelling and computational algorithms, in close interaction with experiments, 
led to the discovery of the extraordinary properties of
two-dimensional materials, such as high carrier mobility, Dirac cone dispersion and bright exciton luminescence, and inspired new device design paradigms. 
This review aims to describe the computational techniques used to simulate and predict the optical, electronic and mechanical properties of two-dimensional materials, and to interpret experimental observations.  
In particular, we discuss in detail the particular challenges arising in the simulation of two-dimensional constrained fermions and quasiparticles, and we offer our perspective on the future directions in this field.
\end{abstract}

\maketitle
\def\thefootnote{$\dagger$}\footnotetext{These authors contributed equally to this work}

\section{Introduction}
 The behaviour of quantum particles, either bosons or fermions, is affected by the dimensionality of the space accessible to them. 
Developments in nanotechnology and nanofabrication over the last few years made it possible to use dimensionality as a parameter, which can be harnessed by precise control of the atomic structure.  In particular, the fabrication of  two-dimensional (2D) materials -- materials where electrons, phonons or other particles are constrained to a 2D manifold -- have revealed a plethora of new phenomena that significantly  enriched our knowledge in condensed matter physics. Starting from the synthesis of graphene, the epitome of 2D materials, several novel layered crystal families, such as metallic  and semiconducting dichalcogenides, silicene, germanene, and phosphorene,
have been discovered. These findings are a potentially disruptive innovation in optoelectronics, spintronics, electromechanics, energy storage, and thermoelectrics.

To investigate the properties of these novel 2D materials, which are dramatically modified by the reduced screening and quantum  confinement, numerical simulations based on first principles, multiscale and more recently machine learning \cite{zunger2018inverse,ren2020inverse} approaches represent an essential tool to interpret and guide experiments. 2D materials can be sub-nanometer thick and may have only a few
atoms per unit cell, making it possible to model them with great precision. In fact, surfaces
and interfaces are atomically sharp and can be represented accurately by atomistic models. 
Nevertheless, the coexistence of atomic and solid state features of these 2D architectures have pushed the existing theories to their limits and created a quest for new computational modelling strategies to assess accurately and efficiently both ground and excited state properties.

This manuscript thus reviews (addressing, in particular, the experimental reader) state-of-the-art computational methods used to model 2D materials at different levels of theory. 
We discuss the theoretical foundations of the numerical methods used to characterise ground and excited-state properties of 2D materials, their respective range of applicability, and we show a few illustrative examples of the computation of electronic, optical, transport and mechanical properties. 
Finally, we reflect upon future challenges and directions of this rapidly evolving field. 

The article is structured as follows: in section \ref{DFT}, we shortly review density functional theory (DFT) for the electronic ground state; in section \ref{MBPT}, we outline computational approaches based on many-body perturbation theory (MBPT) for electronic excited states; in section \ref{TRANS}, we discuss several methods to carry out charge transport simulations; and finally sections \ref{MECH} and \ref{ferro} are devoted to  computational approaches for the simulation of mechanical properties and ferroelectric phases of 2D materials.

\section{Density functional theory} \label{DFT}
DFT is one of the most used computational methods in materials research and 2D materials in particular. It is implemented in numerous codes and has achieved a high level of reproducibility \cite{Lejaeghereaad3000}. We will briefly introduce the fundamentals of DFT and discuss some of the achievements and challenges specific to 2D and layered materials.

\subsection{Fundamentals}
DFT is rooted on the two Hohenberg--Kohn theorems \cite{Hohenberg-PR-136-B864}.
It is an exact formalism, based on a variational principle, that allows one to find the energy and  electron density $n({\bf r})$ of the ground state $|\Psi\rangle$ of a quantum system.
This is conceptually attractive because, unlike the wavefunction, the charge density is an observable.
However, in practice, the determination of the electron density of real systems relies on approximations to describe the exchange and correlation interactions between electrons \cite{Gillan}. In particular, the exchange interaction, whereby many-body systems must be antisymmetric under exchange, stems from Pauli's exclusion principle and acts upon electrons with the same spins. The latter force leads e.g. to spin alignment in ferromagnets or to the Hund's rule in atoms.

The exchange energy functional is known exactly for the homogeneous electron gas, as a function of the charge density, and this can be used as an approximation in solids \cite{KS65}. 
This approximation, known as local density approximation (LDA), can be remarkably accurate for ground states that deviate tremendously from the 3D homogeneous electron gas. It is even applicable to 2D materials, even though it fails in the true 2D limit \cite{pollack2000evaluating}. The remarkable transferability of the LDA and other exchange functionals beyond LDA \cite{2D_PBE_Constantin} has been one of the reasons for the success of DFT.
To treat magnetic systems, it is possible to split the charge density into up and down spin densities 
$ n_\uparrow({\bf r})+n_\downarrow({\bf r}) = n({\bf r}).$ This is designated local spin density-approximation (LSDA).
Nonetheless, the LDA does not provide an adequate treatment of materials that contain transition metal or rare-earth metal ions with partially field $d$ or $f$ shells ($strongly$ correlated systems). Among these, Mott insulator and ordered systems can be described with a popular DFT approach, the LDA+$U$ method \cite{Anisimov_PhysRevB.44.943}. The local Coulomb repulsive ("Hubbard") $U$ for the $d$ and $f$ states  is added to the LDA functional. The Hubbard $U$ is often taken as parameter but it can be also deduced from first principles calculations \cite{Cococcio_PhysRevB.71.035105,Ferdi_cRPAPhysRevB.57.4364,LDA_HF_PhysRevB.76.155123}. However, LDA$+U$ is not able to describe the strongly correlated systems and the Mott-Hubbard metal insulator transition.\cite{Rohringer_RMP.90.025003} 
In this regard, LDA and LDA$+U$ can be considered as the limiting cases of the Dynamical Mean Field Theory (DMFT) (Section \ref{DMFT_section})(see REF. \cite{Held_DMFT_LDAU} ). Moreover, LDA$+U$ may also be regarded as an approximation to the static GWA  (see \ref{GW_section}) \cite{Anisimov_1997}. 

 The remaining term of the electron-electron interaction is the correlation term \cite{Gillan,WS}. The effect of this potential becomes visible e.g. in the dissociation of the H$_2$ molecule, where it is responsible for the electrons ending up on different atoms upon {\it adiabatic} bond stretching~\cite{helbig2009exact}.
We notice that in the context of DFT, the correlation energy is defined differently from other fields of computational chemistry and condensed matter physics \cite{gross-in-lair-book, gritsenko-JCP-107-5007}.

\subsection{Correlation in 2D materials\label{correlation}}
In 2D materials, the charge density can be easily changed by back-gating, enabling correlations to be tuned experimentally in an unprecedented way. In graphene, for example, correlation becomes important when the Fermi energy is close to the Dirac point, and the density of free charges is low~\cite{kotov2012electron}. 
For strongly correlated electronic phases, where electrons are no longer weakly interacting, the DFT treatment no longer provides a valid physical picture. 
An example of this DFT failure is the description of the electronic states in bilayer graphene twisted at a small angle, forming moir\'e superlattices. At specific twist angles, the electronic band dispersion close to the Fermi level is nearly flat in the reciprocal space. When the flat bands are partially filled, the resulting electronic states are highly correlated and can display orbital magnetism, superconductivity, or quantised anomalous Hall effect.\cite{cao2018correlated,cao2018unconventional,yankowitz2019tuning,lu2019superconductors,serlin2020intrinsic}. Other correlated systems, such as low-density electron gas phases in graphene, are still object of research.
Accurate numerical treatment of correlation, such as the random phase approximation (RPA), will be examined in sec.~\ref{MBPT}. 

Alternatively, the Reduced Density Matrix Functional Theory (RDMFT), 
which upgrades DFT by using the reduced density matrix rather than just the spatial density,
has proven the ability to deal with strongly correlated systems dominated by static correlations, and in particular to calculate fundamental band gaps of
semiconductors/insulators and in transition metal oxides \cite{PhysRevB.78.201103_sangeeta,PhysRevLett.110.116403_sangeeta}, where DFT fails. 

\subsection{Correlation and van der Waals forces}
\begin{figure*}
    \centering
    \includegraphics[width=12cm]{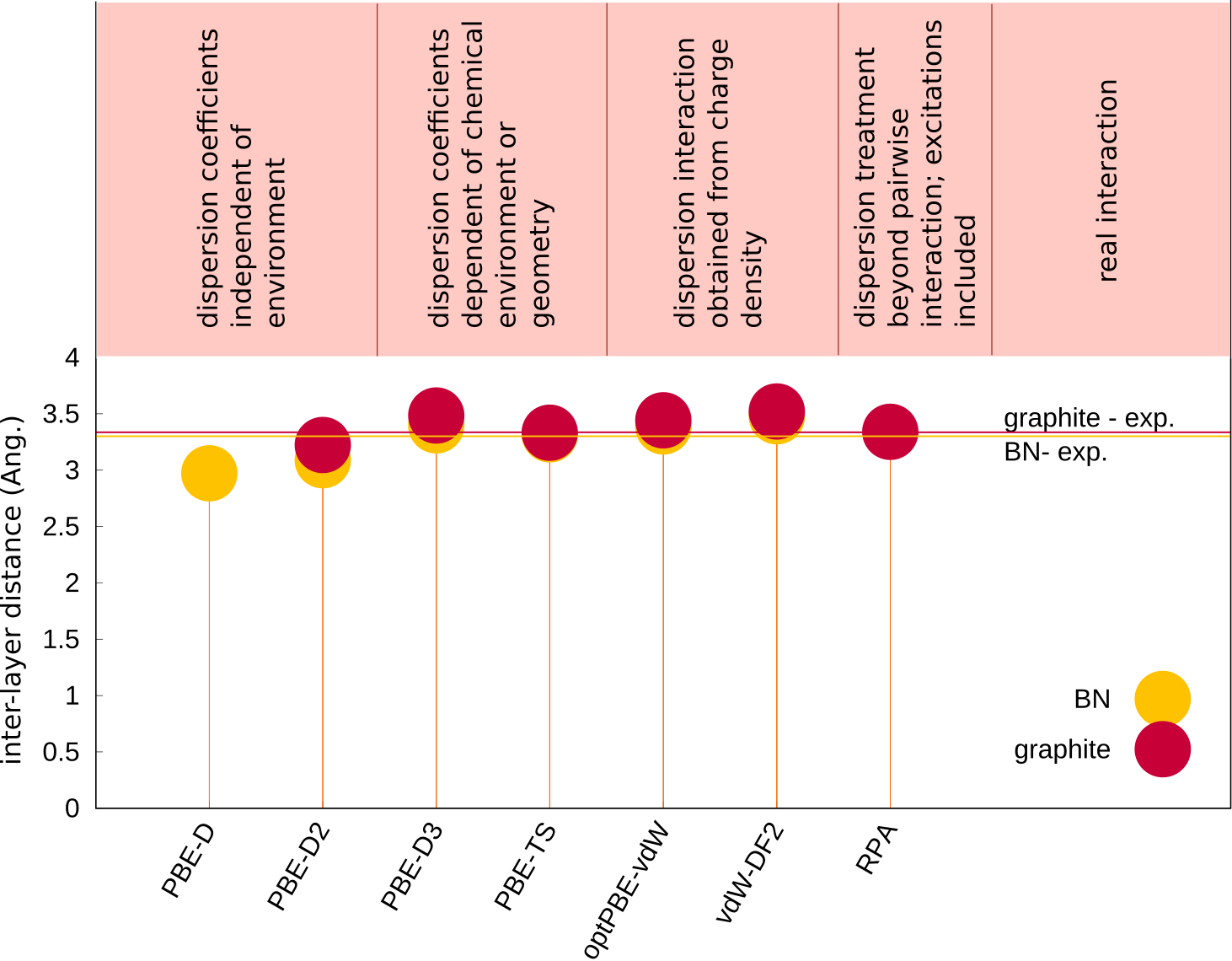}
    \caption{{\bf Van der Waals functionals.} Comparison of the interlayer distance yield by different levels of treatment of the van der Waals interaction for bulk graphite and boron nitride. Experimental reference values at low temperature  are represented by the horizontal lines. Values are from REF.\cite{lebedeva2017comparison}}
    \label{fig:vdw}
\end{figure*}

Correlation is also in part responsible for the van der Waals (vdW) bonding \cite{kawai2016van}.
When two-dimensional layers pair up, a charge density redistribution, arising from induced transient dipoles associated with electronic charge fluctuations, generates an attractive force between layers.
It is known that local or semi-local exchange-correlation functionals developed for 3D solids yield incorrect vdW binding energies and geometries. However, a correct description of vdW systems can be achieved with low computational overhead by adding a semi-empirical dispersion potential, such a pair-wise force field optimized for several popular DFT functionals, to the conventional Kohn-Sham DFT energy (see Fig.~\ref{fig:vdw})~\cite{klimevs2012perspective,grimme-review,johnson-review,burns-review,bjorkman2012van,lebedeva2017comparison}. Correlation energy expression to obtain dispersion-corrected functionals to deal with weak interactions can be derived also using the Adiabatic-Connection Fluctuation–Dissipation theorem approach (see for example ref. \cite{ACFD_thyghess}) within the RPA approximation. These functionals are able to determine the relative thermodynamic stability of different vdW layered superlattices.
Conversely, one can argue that the discovery of 2D materials has contributed to this important methodological development.

\subsection{Choice of basis set for 2D materials\label{basis}}
The majority of DFT codes expand the Kohn-Sham eigenfunctions into a plane wave (PW) basis set~\cite{giannozzi2017advanced,kresse1996efficient}, owing to the fact that in principle the accuracy can be 
indefinitely increased by simply increasing the energy cutoff ($E_{\rm cutoff}$). 
However, the presence of vacuum between 2D layer replicas and inter-layer spacing raise the computational cost, as the number of PWs increases rapidly at a given $E_{\rm cutoff}$. We will come back to this replica problem in sec. \ref{MBPT}.

Owing to the hybrid character of 2D materials, the ideal basis set to deal with layered systems, however, is one in which its functions are periodic in-plane while atomic-like along the perpendicular direction \cite{trevisanutto2016ab,PhysRevB.95.165130_namdo}.

\subsection{Electronic structure}
 The electronic band structure can be calculated (with little formal justification) from the Kohn-Sham orbital energies $\epsilon_l({\bf k})$ as a function of the crystal momentum ${\bf k}$, and the Fermi energy can be obtained from the normalisation constraint.
However, we notice that the single electron energies $\epsilon_l({\bf k})$ describe the dispersion for the auxiliary system obtained by creating an effective potential that is felt by non-interacting electrons. Thus, in metallic systems where e--e interactions dominate, far from the Landau Fermi liquid picture, the one-electron energies are not a good representation of the many-body electron system. Still, in some instances, they may be used due to their conceptual simplicity.\\

\subsection{Limitations}

DFT is, at its origin, a theory of the ground state. 
Thus, even though transition energies, such as the bandgap, have been calculated using the Kohn-Sham states, this approach is prone to very-well known failures, such as the bandgap underestimation when a local density approximation is used.\cite{Hohenberg-PR-136-B864}
Generalised Kohn-Sham schemes addressing both the physical meaning of the Kohn-Sham states as energies and their accuracy have been recently developed~\cite{seidlPhysRevB.53.3764,perdew2017understanding}.

Moreover, the solution of the Kohn-Sham equations using an auxiliary non-interacting gas is not applicable for strongly interacting electronic phases and electronic excitations, as those considered in section~\ref{correlation}. 
More accurate methods for reckoning the electronic fundamental band gap and optical spectra of 2D materials will be discussed in Section~\ref{MBPT}.
One of the alternative methods for the treatment of correlated electron phases will be considered in section~\ref{DMFT_section}.

\begin{figure}
   \includegraphics[width=8cm]{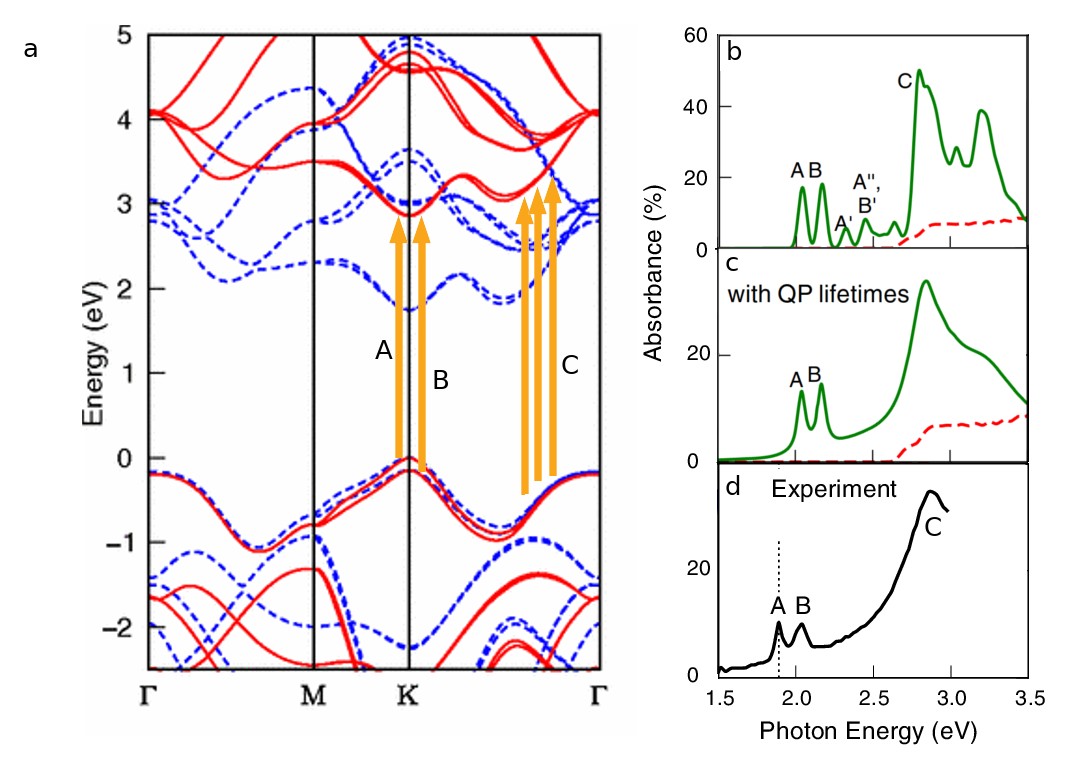}
  \caption{{\bf Origin of the excitons in the optical absorption spectra of monolayer MoS$_2$}. {\bf a}$|$ LDA (dashed blue curve) and GW (solid red curve) band structure. Arrows indicate the optical transitions.
  {\bf b}$|$ Absorption spectra without (dashed red curve) and with (solid green curve) electron-hole interactions, calculated by solving the BSE.  {\bf c}$|$ Same as {\bf b} but using an {\it ab-initio} broadening based on the electron-phonon interactions, ie. with quasi-particle lifetimes.
  Panel {\bf a} is adapted from REF.~\cite{qiu-PhysRevLett.111.216805}, Copyright (2013) by The American Physical Society. Panels {\bf b-c} are from REF.~\cite{PhysRevLett.115.119901}, Copyright (2015) by The American Physical Society. Panel {\bf d} is from REF.~\cite{mak-PhysRevLett.105.136805}, Copyright (2010) by The American Physical Society.
\label{fig-excitons}
}
\end{figure}

\section{Ab-initio Many-body Perturbation Theory in 2D Materials}\label{MBPT}

\subsection{The GW approximation}\label{GW_section}

Excited-state properties and the interpretation of electron spectroscopy experimental measurements in condensed matter physics rely on the concept of quasiparticle (QP), which is represented as a fundamental excitation of bare particle plus its polarization cloud. QPs are not eigenstates of the Hamiltonian  and thus acquire finite life-times. In this subsection, we will limit our discussion to particular QPs, the electronic polarons (as originally defined by L. Hedin \cite{Hedin_pes}), which can be imagined as electrons (or holes) dressed by their electronic clouds.

Within the \textit{ab-initio} many-body perturbation theory (Ai-MBPT) framework for electronic structure calculations, the central quantity is the one-particle Green's function $G$ (at zero temperature), which is defined as the inverse of the ground state expectation value of the interacting Hamiltonian of the $N$ electron system ($\ket{N}$). Using this definition, one can write the Dyson equation as follows:
  \be
  G^{-1}(\br_1,\br_2,\omega)\equiv G^{-1}_{0}(\br_1,\br_2,\omega)- \Sigma(\br_1,\br_2,\omega)
  \label{eq:1Green}
  \ee
where $G^{-1}_{0}(\br_1,\br_2,\omega)$
is the inverse of the non-interacting Green's function and $\Sigma(\br_1,\br_2,\omega)$ is the self-energy. $\Sigma(\br_1,\br_2,\omega)$ is a non--Hermitian, energy--dependent, and non-local quantity, which ``dresses'' the bare particle with the whole many-body electron interaction cloud. 
To describe some important features of Ai-MBPT, it is useful to write $G$, 
through the Lehmann spectral representation \cite{RevModPhys.74.601_onida,Hedin_pes,HEDIN19701}, in terms
of the $f_s$, the transition amplitudes from the $N$-electron system state $\ket{N}$ to the excited state $s$ of the $N\pm1$ system $\ket{N\pm 1;s}$  In Ai-MBPT, the quantum states $\ket{N}$ to $\ket{N\pm 1;s}$ are determined from either the non-interacting DFT or a tight-binding (TB) approximation (see sec.~\ref{Tbind}). 
The QP energies $E^{QP}_{s}$ and the amplitudes $f_s$ are obtained from the homogeneous equation \eqref{eq:1Green}:
  \be
  \left(E^{QP}_{s}-h-\Sigma(\omega)\right)f_s(\br)=0
  \label{eq:Hom_eq_gw}
  \ee
where the real part of the eigen-energies $E^{QP}_{s}$  delivers the QP band structures, while the imaginary part represents the QP damping (by means of the imaginary part of $\Sigma$). Both $f_s$ and $E^{QP}_{s}$ are directly related to the the spectral function of particles and, therefore to the intrinsic losses in photo-emission spectroscopy (PES) and inverse PES.

GW is the most successful and popular approximation to the self--energy in Ai-MBPT. With respect to DFT calculations, GW improves the agreement with the experimental data of the simulated electronic band structure  \cite{RevModPhys.74.601_onida}, and reads (the symbol ``*'' stands for the convolution product):
  \be 
    \Sigma (\br_1,\br_2,\omega) = \frac{i}{2\pi} G_0(\br_1,\br_2,\omega)*W(\br_1,\br_2,\omega)e^{-i\eta \omega}
  \label{eq:GW}
  \ee
where $W$ is the screened Coulomb interaction that is related to the \textit{bare} Coulomb potential $w$ via the dielectric function $\epsilon$: 
   \be
    W (\br_1,\br_2,\omega)=\int d \br_3\epsilon^{-1}(\br_1,\br_3,\omega)w(\br_3,\br_2,\omega)
  \label{eq:epsilon-W}
  \ee
$\epsilon(\br_1,\br_3,\omega)$ measures the tendency of a medium to be polarized under the action of an external electromagnetic field, it is related to the energy loss spectrum \cite{PhysRevB.96.235414_Nazarov2} and energy loss in charge transport Monte Carlo (see sec. \ref{TRANS}).

We notice that the GW self-energy is the HF exchange potential in Eq.~\eqref{eq:GW} if the screened Coulomb potential $W$ is replaced by the bare Coulomb potential $w$.
  
\subsection{The Bethe--Salpeter equation}\label{BSE}

Besides one-particle properties, two-particle properties also play a prominent role in the study of excited-state features of 2D materials, namely excitonic effects in absorption spectra. A typical example is the electron-hole (e-h) QP: the exciton. Similarly to QPs theory, the two-body Green's function $L$ is determined by solving the Bethe--Salpeter equation (BSE):
   \be 
   \begin{split}
   L^{-1}_{vc,v'c'}(\omega)&\equiv H^{\rm BSE}_{vc,v'c'}(\omega)\\ &=(E^{\rm QP}_{c}-E^{\rm QP}_{v})\delta_{v,v'}\delta_{c,c'}+I^{eh}_{cv,c'v'}(\omega)  
   \end{split}
  \label{eq:BSE}
  \ee
where $v$ and $c$ are valence and conduction one-particle states, $H^{\rm BSE}_{vc,v'c'}(\omega)\equiv \bra{vc}\hat{H}\ket{v'c'}$ is the effective two-body Hamiltonian, and $I^{\rm eh}_{cv,c'v'}$ is the kernel of this Dyson-type equation, which includes the e-h interactions. 
In Ai-MBPT, an important assumption for semiconductors and insulators is to approximate the kernel $I$ with the static screened potential $I\approx W$. 

In bulk materials, these properties are calculated from the macroscopic dielectric function $\epsilon_{M}(\omega)$ which is related to the imaginary part of $L$ ($q$ is a reciprocal lattice vector) through:
  \be
    \epsilon_{M}(\omega)=\lim_{q\to0}(1-v(q)\Im L(q,\omega))
  \label{eq:mac_epsilon}
  \ee
where the long wavelength limit of the interaction Coulomb potential $v$ must be considered.
In 2D materials the macroscopic dielectric function reduces to 1 and, to determine the optical properties, the  longitudinal in-plane polarisability has to be considered \cite{trevisanutto2016ab}:
  \be
    \chi_P(\omega)\approx\lim_{q\to 0} \frac{1}{q^2} L(q,\omega)
  \label{eq:mac_epsilon_polar}
  \ee
Alternatively, one can obtain Eq. \eqref{eq:mac_epsilon} by averaging $W$ over the material rather than the simulation supercell \cite{PhysRevB.88.245309_thyge1}.
  
Figure~\ref{fig-excitons} shows how GW+BSE can be used to interpret the exciton features in MoS$_2$, including the optical absorption threshold, exciton binding energy, and spin and momentum-related selection rules.
  
\subsection{The screened potential $W$ in 2D materials}

The screened potential $W$ enters in both these approaches, and is by far the most difficult to estimate amongst the many-body terms.
It is calculated starting from the non interacting polarization function $P_0$, which is determined from the ground state energy and wave function (DFT or TB). The minimal basis set necessary to obtain accurate results comparable with experiments is usually very large. 
This issue is increased in quasi-2D materials, where the crystalline periodicity is present only in the planar directions whereas in the $z$ direction (normal to the plane) the wave functions have a polyatomic molecular character. This hybrid (both crystalline and molecular) behaviour induces, on the plane of the layers, electric field fluctuations that are not screened at large distances: the e-e and e-h interactions are thus much stronger than in 3D materials, producing excitons with notably high binding energy \cite{qiu2013optical,ugeda_giant_2014,PhysRevB.89.235319_liyang_bp,acs.nanolett.5b00160_bradley,liu_evolution_2014, doi:10.1021/acs.nanolett.5b01251_genome,PhysRevB.94.155428_lidia, guo_exchange-driven_2019,arora_interlayer_2017, zheng_acs_nano,doi:10.1021/nl401544y_maurizia} and, in presence of defects, extremely stable \cite{zheng_acs_nano}. 

\indent Due to their high binding energy, excitons in quasi-2D materials are stable and localized. Qubits \cite{maragkou_dark_2015}, {single-photon light emitter devices using the self-trapping effects \cite{aharonovich_solid-state_2016}, excitonic devices or cavity polaritons \cite{unuchek_room-temperature_2018}, and valleytronics \cite{schaibley_valleytronics_2016} are some of the potential applications exploiting these peculiar many-body effects in 2D materials 
\cite{qiu2013optical,ugeda_giant_2014,PhysRevB.89.235319_liyang_bp,acs.nanolett.5b00160_bradley,liu_evolution_2014, doi:10.1021/acs.nanolett.5b01251_genome,PhysRevB.94.155428_lidia, guo_exchange-driven_2019,arora_interlayer_2017, zheng_acs_nano,doi:10.1021/nl401544y_maurizia}.

By using periodic Ai-MBPT plane-wave codes, quasi-2D materials calculations present the replica issue previously mentioned (see sec. \ref{basis}). Owing to the ineffective Coulomb interaction screening of the bare $w$, the required distance between the replicas can be very large, making calculations unfeasible, as the mesh grid for reciprocal space integration has to be increased.
To overcome this problem, the Coulomb interaction is truncated beyond a certain cutoff distance in the $z$-direction  \cite{PhysRevB.73.205119_rozzi_cutoff,PhysRevB.80.033102_castro_cutoff,PhysRevB.88.245309_thyge1,PhysRevB.92.245123}. 
A slightly different approach~\cite{PhysRevLett.104.226804_cudazzo,PhysRevB.84.085406_cudazzo} consists in replacing $w$ by a Keldysh-like potential to simulate the in-plane screening of a 2D layer in vacuum, 
  \be
  W_{2D}(q)=\frac{2\pi e}{|q|(1+\alpha_{2D}q)}
  \ee
where $e$ is the charge and $q$ is a reciprocal lattice vector in the plane.
This potential is often used in connection with TB calculations \cite{PhysRevB.92.235432_Pedersen_2015,PhysRevB.97.205409_ridolfi}.

\indent Finally, 2D materials are often encapsulated or lying over substrates. Several works were devoted to deal with substrate effects on $W$  \cite{Trolle2017,PhysRevB.92.235432_Pedersen_2015,PhysRevB.90.075429_Rodin,PhysRevB.82.205127_GdW,PhysRevB.92.245123}.

\subsection{Second, third, and higher harmonic generations}
  
The strong optical response of 2D materials to external electromagnetic fields generates peculiar effects also in non-linear regime, giving rise to an optical response at multiples of the exciting frequency $\omega$. Graphene and TMD semiconductors show non-trivial topological characteristics  \cite{Xu2014,Liu2016b,schaibley_valleytronics_2016}, which have been investigated with higher harmonic generation calculations \cite{luppi2016}. To deal with these physical mechanisms, two different main approaches are developed in the context of Ai-MBPT: the direct integration of the time-dependent macroscopic polarization and the perturbative approach for the non-linear susceptibilities. 
   
The macroscopic polarization $P(t)$ is the key observable of interest for optical properties. This can be expressed in terms of one-particle reduced density matrix as:
  \be
    \bm{P}(t) = e \int \!\br \hat{\rho}(\br,t) d\br
    = \frac{e}{A} \sum_{mn\bk} \br_{mn\bk} \, \rho_{mn\bk}(t),
  \label{eq:P-def-rho}
  \ee
where $e$ is the charge of the electron, and $A$ is the 2D crystal area. The time-dependent density matrix is  defined as $\rho_{mnk}(t) \equiv \langle \hat{a}^\dagger_{mk}(t) \hat{a}_{nk}(t)\rangle$ by the creation (and destruction) operators $\hat{a^{\dagger}}$ ($\hat{a}$). In the length gauge \cite{PhysRevB.52.14636_aversa_sipe}, the dipole matrix elements $\br_{mnk} \equiv \bra{mk}\hat{\bm{r}}\ket{nk}$ are separated in two components,

the  intraband (or Berry Connection \cite{RevModPhys.66.899_Resta}) $\bm{r}^i$ and the interband dipole operators $\bm{r}^e$ where the indices $m$ and $n$ are the bands. 

\indent In the high harmonic generation optical responses, the e-h interactions are dominant.
An Ai-MBPT description is provided by the time-dependent (TD) BSE \cite{Attaccalite2011}. The TD-BSE is originally derived from the Kadanoff-Baym equation and it can be related to the Linblad master equation for open quantum systems \cite{breuer_theory_2007,PhysRevB.97.144302_wismer}. If the interaction with the external electric field $\bm{F}$ is written in the length gauge  ($H_{int}=e\bm{F}\cdot\bm{r}$)  TD-BSE reads
\begin{equation}
  i\hbar\frac{\partial}{\partial t}G_{\bk}(t)
  = \Bigl[h_{\bk}+e\bm{F}(t)\cdot \bm{r}+\Sigma_{\bk}[G_{\bk}(t)], 
  G_{\bk}(t) \Bigr],
  \label{eq:dGdt-1}
\end{equation}
where  the self-energy $\Sigma$ is introduced either in the HF approximation or in the static GW self-energy (COHSEX), and $[.,.]$ is the commutator. In absence of the self-energy (and then, the e-h interactions), the \textit{``semiconductor Bloch equations''} (SBE) is retrieved from the TD-BSE when considering the connection between the Green's function and the density matrix: $\bm{G}_{nm\bk}(t)\equiv \bm{G}_{nm\bk}(t,t)=i\rho_{nm\bk}(t)$. In the strong-field resonant dynamics and attosecond optoelectronics, the SBE is currently implemented to calculate the band population (diagonal terms) and coherences (the off-diagonal terms, see subsection \ref{Trion_biexciton_ss})  \cite{heide_attosecond-fast_2020,sederberg_attosecond_2020,PhysRevLett.116.197401_wismer}.
Analogous formulas for the density evolution can be retrieved within the time-dependent DFT (TD-DFT) framework, leading to equations analogous to the SBE and TD-BSE (see \cite{luppi2016} for an exhaustive review on the topic).\\
\indent From Eq. \eqref{eq:dGdt-1} the non linear responses such as the second, third, and higher harmonic generations can be estimated with the perturbative derivation of the harmonic generation response functions    \be
      P=\chi^{(1)}F +\chi^{(2)}FF + \chi^{(3)}FFF+...
      \label{eq:Polarization_higher_orders}
   \ee
where $\chi^{(1)}$, $\chi^{(2)}$, $\chi^{(3)}$ are, respectively the macroscopic first, second and, third, order susceptibilities. These high order responses can be analytically derived by expanding the Eq. \eqref{eq:dGdt-1} in terms of one-particle Green's function $\bm{G}_{nm\bk}(\omega)$, $\bm{G}_{nm\bk}(2\omega)$, $\bm{G}_{nm\bk}(3\omega)$ \cite{PhysRevB.92.235432_Pedersen_2015}. 

Alternatively, macroscopic polarization $P$ can be obtained through Eq.~\eqref{eq:P-def-rho} and with the Green's function time-integration scheme of the Eq.\ref{eq:dGdt-1}. The linear response (i.e. the BSE \eqref{eq:BSE}) and the higher order generation responses $\chi^{(i)}$ are retrieved by numerical Fourier transform of $P(\omega)$  \cite{Takimoto_2007,Attaccalite2011}. This approach allows for high harmonic generation. Nevertheless, it requires a particular care when extracting the non-linear susceptibilities: the $P(t)$ can be split in the driven (whom we are in interested in ) $P_{driven}(t)$ and and a decaying transient $P_{trans}(t)$. The latter could conceal the higher nonlinear harmonic generations \cite{2020_Ridolfi_PhysRevB.102.245110}. \\
\indent
Ai-MBPT calculations of the optical response of boron nitride show that excitons also dominate the  nonlinear optical properties \cite{PhysRevB.92.235432_Pedersen_2015}.
Moreover, crystal local field effects, whose importance was already displayed for bulk and surfaces \cite{luppi2016}, dramatically shape the SHG and THG spectra in MoS$_2$ and h-BN monolayers \cite{Attaccalite2014}.
Finally, differently from the linear response regime, the interband and intraband transitions cannot be separated as shown in the SHG and THG calculations in the spectrum of bilayer graphene \cite{PhysRevB.98.205420_fabio_bilayer}, and in the THG calculations of semiconductor BP \cite{PhysRevB.97.035431_fabio_bp}, MoS$_2$ and h-BN  \cite{PhysRevB.92.235432_Pedersen_2015} SLs. 
Real-time simulations have recently been used to better determine the model used to extract the SHG coefficient from the experimental data, such as in the case of the 2D monochalcogenides GaSe and InSe \cite{PhysRevMaterials.3.074003_gruning}.
 
\subsection{Trions, Exciton--polarons, and Biexciton formations}\label{Trion_biexciton_ss}
 
Controlling the optical response of the material by doping the substrate or by  gating is a possibility in 2D materials that has had no parallel in traditional optical materials. Recently, 2D TMDs optical absorption experiments \cite{sidler_fermi_2017,mak_tightly_2013} have shown that, in the presence of free carriers, the most prominent excitonic features split into two distinctive peaks. The second peak was initially attributed to trions, i.e. the three--body QP characterised by a bound state of an exciton plus an hole or electron, and later to the dressed excitons, the \textit{exciton-polarons}, that is the many--body generalization of the trion bound and unbound states. It has been theoretically shown that at low doping, the ground state corresponds to the trion and becomes an exciton-polaron at higher doping. \cite{exciton_pol_McDonald,PhysRevX.10.021011_pol_polaritons,PhysRevB.98.235203_combescot}.
From an \textit{ab-initio} point of view, this problem was tackled by extending the Ai-MBPT to include this  three-body effective Hamiltonian (e.g. in carbon nanotubes \cite{PhysRevLett.116.196804_rohlfing,PhysRevLett.123.259902_erratum_rohlfing}), or to describe the optical spectrum lineshape (e.g. of MoS$_2$ \cite{trion_MoS2_nature}). In a (two-electrons) plus hole system, $eeh$ (the electron and two holes, $hhe$, is analogous) the effective Hamiltonian is approximated as
an extension of the BSE \eqref{eq:BSE}: the particle-particle interactions are described by the single BSE (an then mediated by the Screened many--body interaction $W$) correlations with the third particle non interacting. The three-body correlation kernel term is disregarded. 
This Hamiltonian, with a combination of GWA and Configuration Interaction (CI) approaches, has been applied to the case of WS$_2$ and MoS$_2$ monolayers \cite{PhysRevB.100.201403_torche_trions}.\\
\indent These calculations provide quantitative binding energies of the trion resonances both in a free-standing layer and in a more realistic case with a substrate (or encapsulation) that enhances the environmental screening. More importantly, the analysis of transitions has shown the nature of trions \cite{trion_MoS2_nature} (Fig. \ref{fig-trions}).

Recently, ultrafast pump-probe experiments on monolayer WSe$_2$ showed the biexcitons signatures and a fine structure in excellent agreement with the theory. The dynamics-controlled truncation theory for biexcitons appear in the THG susceptibility \cite{Nature_biexciton}. The biexciton spectrum was modeled via an SBE--like equation of motion for the four--particle correlation function:
\begin{widetext}
\begin{equation}
B_{k,k'}^{c'v'cv}(q,t)\equiv \braket{\hat{a}^\dagger_{c'(-k'-q)}\hat{a}_{v'k'}\hat{a}^\dagger_{c(-k)}\hat{a}_{v'(k+q)}}(t) -\braket{\hat{a}^\dagger_{c'(-k'-q)}\hat{a}_{v'k'}}(t)\braket{\hat{a}^\dagger_{c(-k)}\hat{a}_{nv(k+q)}}(t)
\label{fig-4correlationFunction}
\end{equation}
\end{widetext}
where the interaction Hamiltonian is described as the extension of trion and two-body BSE Hamiltonian for the 4-particle $B_{k,k'}^{c'v'cv}$ but in the HF approximation. 

\subsection{Dark and bright exciton formation, photolomuniscence  and biexciton fine structures in monolayer TMDs.}

Besides bright excitons, TMDs exhibit dark excitons that modify the dynamics, the coherence lifetime and, at last, the photo--luminescence. In a series of articles  \cite{Steinhoff_2016_2dmaterials,Selig_2018_2dmaterials,nanoletter_dm1,steinhoff2014influence}, the SBE-like equations for the excitonic density matrices have been extended to include phonon and photon scattering events and describe the exciton dynamics. To summarize, the exciton band structure is calculated from DFT and GW calculations. The Heisenberg equation of motion for the coherent exciton polarization $P_q(t)\equiv\sum_k\psi^*_k)\hat{a}^\dagger_{c(k+q)} \hat{a}_{v(k-q)}(t)$ ($\psi_k$ the exciton wave-function) is derived. The total Hamiltonian in the SBEs includes the ``free" exciton Hamiltonian, incoherent (4-particle) exciton density formation $N_Q(t)\equiv\sum_{k,k'}\braket{\hat{a}^\dagger_{c(k-\alpha q)}\hat{a}_{v(k+\beta q)}\hat{a}^\dagger_{v(k'+\beta q)} \hat{a}_{c(k'-\alpha q)}}(t) $ through the phonon-exciton coupling, and photon-exciton interactions in the low excitation regime. On the same reference, the equation of motion for the incoherent exciton occupation density $N_Q$ is written including in the Hamiltonian the non-radiative decay of $P_Q$, phonon scattering processes (in and out), and the spontaneous photon emission.  
Among the main results, the theory predicted that a carrier relaxation in TMDs occurs on 50 fs time scale, an order of magnitude faster than in quantum wells whereas the incoherent part of the photoluminescence was estimated to decay on a timescale of few tens of nanoseconds.

\begin{figure*}[hbt!]
\centering
\includegraphics[width=0.9\textwidth]{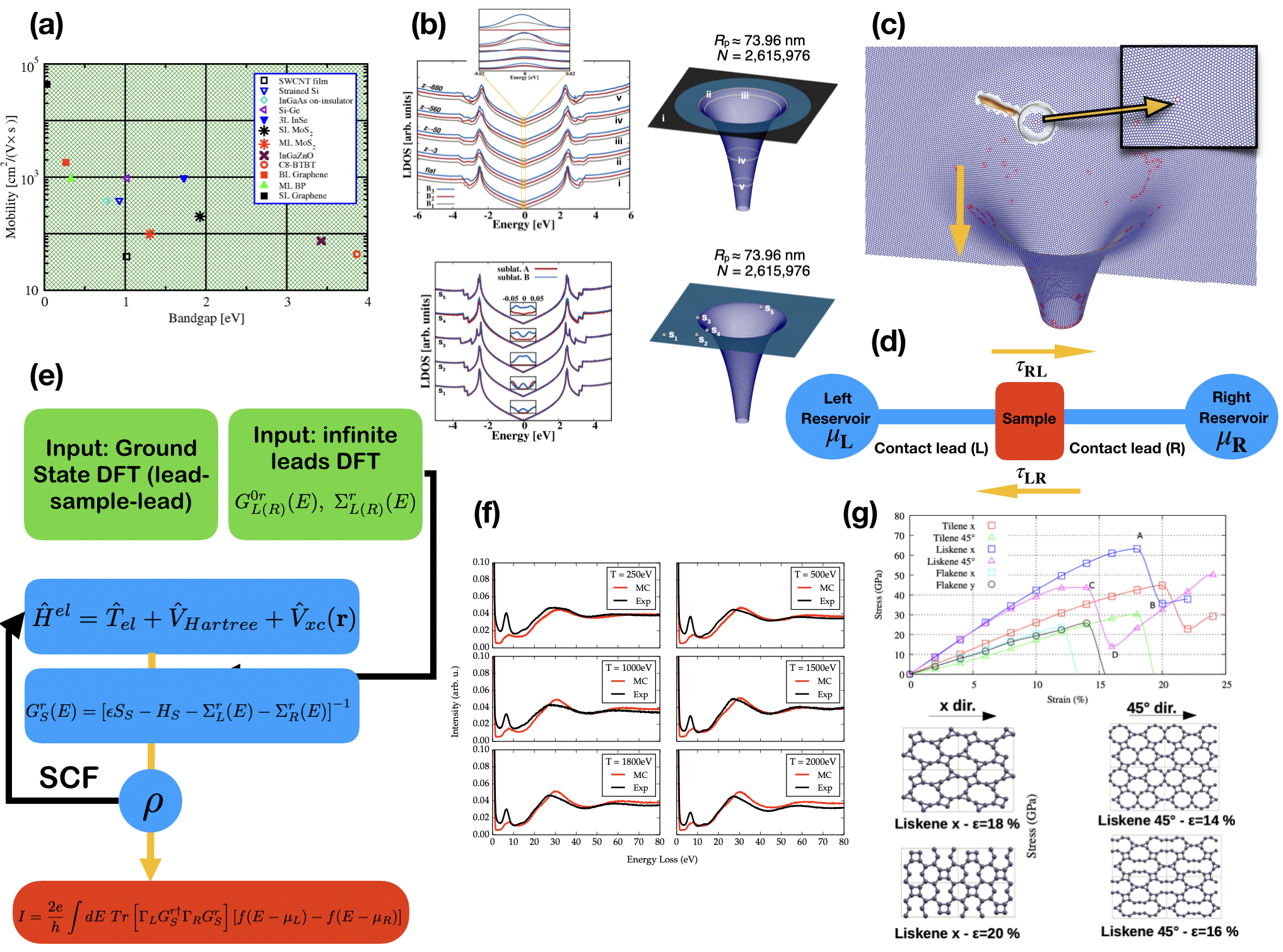}
\caption{{\bf Transport and mechanical properties of 2D materials}. {\bf a}$|$ Room-temperature mobility vs. bandgap of selected 2D materials compared with those of 3D semiconductors. 
ML, SL, 3L stand for few-layer, single-layer and tri-layer, respectively, and BP stands for black phosphorus  \cite{doi:10.1146/annurev-physchem-050317-021353}.
{\bf b}$|$ Up panel: local density of states (LDOS) symmetry breaking due to curvature effects in a pseudosphere graphene membrane with $N= 2,615,976$ carbon atoms, evaluated using a TB approach. Bottom: LDOS projected over the two nonequivalent graphene sublattices A and B near the Fermi energy. {\bf c}$|$ Graphene membrane shaped as a Beltrami's pseudosphere. 
{\bf d}$|$ Set-up of a typical Landauer-B\"uttiker (LB) transport simulation. {\bf e}$|$ Logic flow of the LB {\it ab-initio} approach.  {\bf f}$|$ Reflection electron energy loss spectra (REELS) of a highly oriented pyrolytic graphite (HOPG) sample for several primary beam kinetic energies simulated by MC. {\bf g}$|$ Stress--strain curves of liskene, tilene, and flakene along the different strain directions from first-principles simulations.
Panels {\bf b-c} are republished with permission of IOP, from  REF.\cite{morresi2020exploring}; permission conveyed through Copyright Clearance Center, Inc.
Panel {\bf f} is reprinted with permission from from REF.\cite{azzolini2018anisotropic}. Copyright 2018 American Chemical Society.
Panel {\bf g} is reprinted from REF.\cite{morresi2020structural} Copyright (2020), with permission from Elsevier.}
\label{fig:mobility}
\end{figure*}

\begin{figure*}
   \includegraphics[width=12cm]{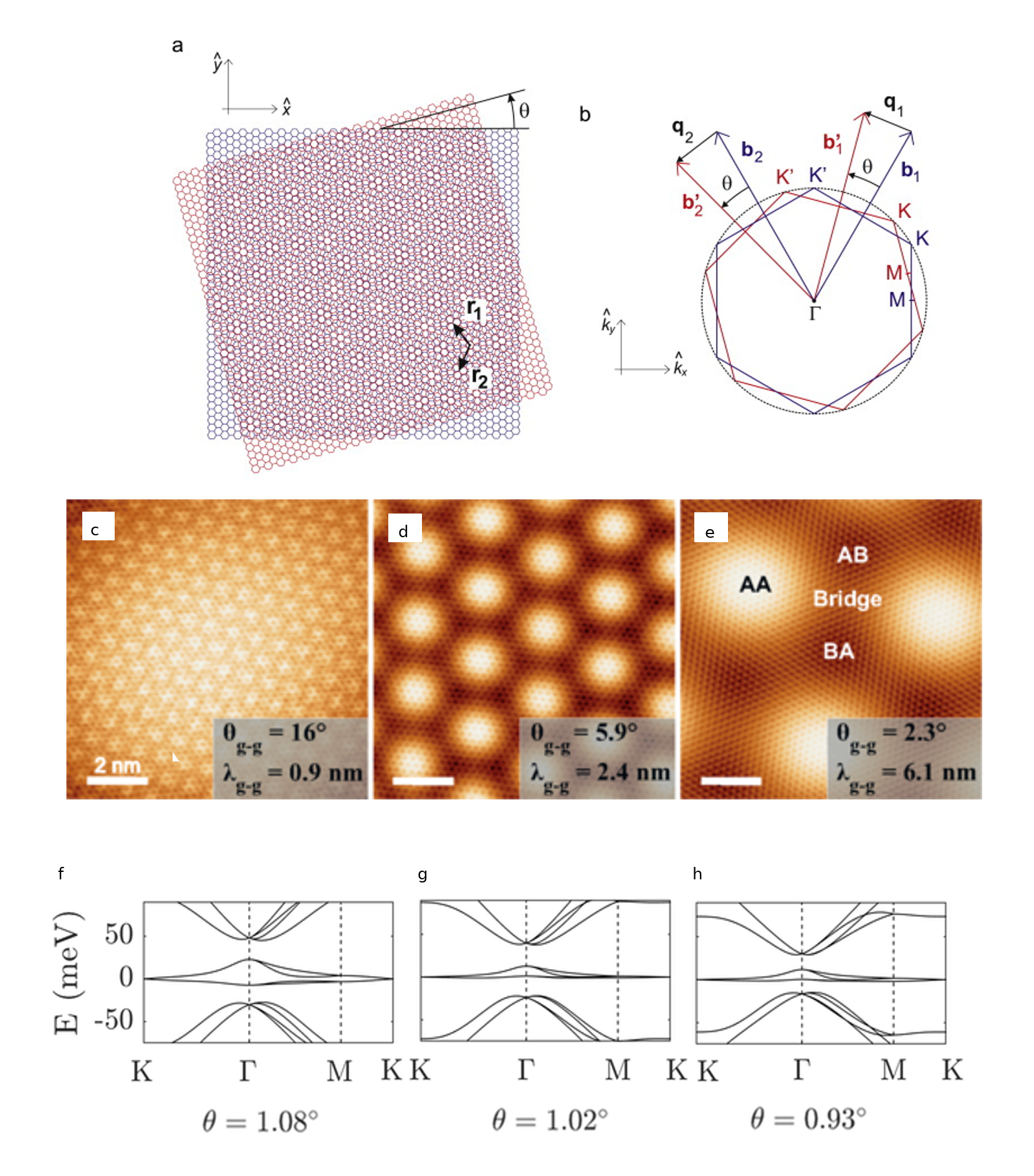}
  \caption{{\bf Tight-binding bandstructure of twisted bilayer graphene.}
  {\bf a}$|$ Moir\'{e} lattice in real space and {\bf b}$|$ its reciprocal space. Primitive cells for the angle ($\theta$) of interest have thousands of atoms and are often too large to be modelled by DFT. {\bf c-e}$|$ Scanning tunneling microscopy images at different angles ($\theta_{\rm g-g}$) and related periodicity $\lambda_{\rm g-g}$. The images are for twisted bilayer graphene over BN, but the BN is not visible.
  Band structures for fully relaxed twisted bilayer graphene obtained with a tight-binding model, for angles slightly above (1.08 $^\circ$), very close (1.02$^\circ$), and slightly below (0.93$^\circ$) the `magic' angle at which the bands close to the Fermi level become flat.
  Panels {\bf a-b} are reprinted from REF.~\cite{carozo2011raman}, Copyright 2011 American Chemical Society. Panels {\bf c-e} are from REF.~\cite{wongPhysRevB.92.155409}, Copyright (2015) by The American Physical Society. Panels {\bf f-h} are reprinted from Ref.~\cite{carr2019exact}, licensed under CC BY 4.0.
\label{fig-twisted}
}
\end{figure*}

\begin{figure}
   \includegraphics[width=8cm]{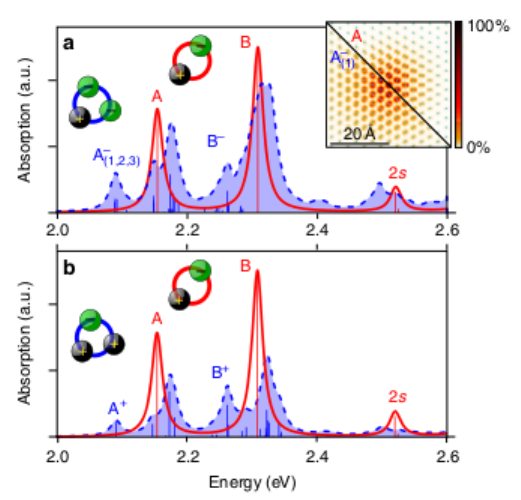}
  \caption{{\bf MoS$_2$ monolayer spectrum from  Ai-MBP} showing excitonic spectrum (red) and trion contribution (blue). {\bf a}$|$ Negative trions and {\bf b}$|$ positive
trions exhibit resonant states (close to the A and B exciton). The A$^{-}$
trion is split into three separate peaks, labeled A$^{(-)}_{(1,2,3)}$.
  From REF.~\cite{trion_MoS2_nature}, licensed under CC BY.}
\label{fig-trions}
\end{figure}

\subsection{Dynamical Mean Field Theory and beyond}\label{DMFT_section}
When dealing with materials with open narrow $d$ and $f$ shells, one of the most successful first-principles method is the Dynamical Mean Field Theory (DMFT). The $strongly$ correlated systems represent a wide solid-state physics field which ranges from quasi-particle renormalizations to magnetic ordering. Many complete reviews about DMFT can be found in the literature (for instance \cite{Georges_RMP.68.13, kotliar_dmft,Held_DMFT_LDAU,Katsenlson_RMP.80.315}). The main DMFT idea is to map the lattice system onto an Anderson impurity model (AIM). This mapping is performed between the AIM Green's function and the local Green's function of the site plus its local Self Energy $\Sigma^{loc}$ which includes the unknown hybridization term (the mean-field). The hybridization term and the $\Sigma^{loc}$ are determined with a self-consistent procedure. This mapping becomes exact in the limit of infinite spatial dimensions (or infinite coordination number) when the correlations become local. Nevertheless, in systems $<2$D and in the proximity of phase transitions with divergent correlation lengths, the DMFT approximation can show some limitations because of the non-local correlation terms. For these reasons, new approaches were developed \cite{Rohringer_RMP.90.025003} to extend the theory beyond DMFT  taking into account the non-local correlations. Cluster extensions of the DMFT such are the dynamical cluster approximation (DCA) \cite{Hettler_DCA_PhysRevB.58.R7475} and the cellular DMFT (CDMFT) \cite{Licht_kats_CDMFT_PhysRevB.62.R9283,kotliar_CDMFT_PhysRevLett.87.186401} extend the impurity model to a cluster embedded in a dynamical mean-field theory. Instead, the diagrammatic extensions to the DMFT, such as D$\Gamma$A or QUADRILEX (see \cite{Rohringer_RMP.90.025003}) start from DMFT to estimate the local irreducible vertex correction diagrams  (i.e. that can be split into two parts when cutting two e-h connecting lines). This is performed utilizing a new self-consistent cycle that includes the BSEs for the full and irreducible vertex \ref{BSE}.
In contrast to DCA, the diagrammatic DMFT extensions fulfil the Mermin and Wagner (MW) theorem. These methods have been applied both to the 2D Hubbard model at half-filling and 2D triangular lattice systems \cite{Rohringer_RMP.90.025003}.

$2$D or layered magnetic materials, which will be the subject of section\ref{magnetic} represent a test-bed to both correctly address the Self Energy non-locality and the long-range fluctuations. The diagrammatic DMFT extensions have been successfully employed in the description of the phase of the square 2D Hubbard model at half-filling and triangular lattice systems \cite{Rohringer_RMP.90.025003}.

\section{Quantum transport} \label{TRANS} 

In layered structures, band dispersion and electron transport are both strongly affected by reduced screening and quantum confinement. This results in a wide range of band gaps and charge carrier mobilities (see Fig.~\ref{fig:mobility}(a)).
On the other hand, highly anisotropic in-plane and out-of-plane electrical properties \cite{Zhang_2017} emerge. For example, in bulk MoS$_2$, the in-plane mobility ($\mu_{x,y}$) exceeds the out-of-plane mobility ($\mu_z$) by a factor of $10^3$ \cite{doi:10.1146/annurev-physchem-050317-021353}, essentially due to the vdW interlayer spacing acting as a tunneling barrier (see Fig. \ref{fig:mobility}(a)).
In this section, we consider transport modelling approaches and their applicability in different regimes.
In particular, we describe the full quantum methods to deal with coherent transport, that is the Landauer-B\"uttiker (LB) and Kubo formalisms. These approaches stem from different viewpoints: in the latter the current is the response to an applied electric field, in the former it is the charge accumulation at the sample boundaries due to the current that generates the field \cite{di_ventra_2008}. These two viewpoints are thus complementary, meaning that they swap the cause and effect. Furthermore, we discuss also the semiclassical Boltzmann transport equation (BTE), which can be obtained as a first-order approximation to the full quantum treatment, and the transport Monte Carlo (MC) method, which can be essentially thought as a numerical method to solve the BTE.}

\subsection{The tight-binding approach\label{Tbind}}

Tight-Binding (TB) \cite{PhysRevB.86.075402} is an approximate quantum-mechanical method to calculate the electronic structure of solids, using a basis set of atomic orbitals $\phi_i(\bf{r})$, and it is widely used for transport modelling in 2D materials.
TB assumes that inside a solid, electrons are tightly bound to the atoms, and that their single particle wave functions $|\psi_{\bf{k}n}\rangle$ in the crystal are quasi-atomic.

The TB scheme has been very successful in modelling graphene, and is easily extendable to other hexagonal 2D materials \cite{morresi2020exploring,Taioli_2016}. 
In the case of graphene, the Dirac Hamiltonian of chiral massless particles of quantum electrodynamics (QED) in two dimensions can be obtained by considering only first nearest-neighbor (NN) interactions between $p_z$ orbitals and by expanding linearly in the momentum the matrix elements of the TB Hamiltonian in the proximity of the $K$ (or $K'$) point. This results in a $2\times 2$ Hamiltonian
\be \mathcal{H}|_K=v_F \hat{\sigma} \cdot \bf{p},\ee
where $\hat{\sigma}$ are the Pauli matrices,  and $v_F=\sqrt{3}\gamma_0 a/(2\hbar)$ is the Fermi velocity. The latter resembles the QED spin Hamiltonian where the Fermi velocity plays the role of the light velocity of massless particles. 
By including interactions up to third NN $\pi-\pi^*$ orbitals, it is possible to recover the band asymmetric behaviour far from the Dirac cones.

The TB model is computationally very efficient as it can easily scale up to several million atoms \cite{PhysRevB.86.075402}.
For example, the TB approach was used to calculate the LDOS in 2D graphene pseudospheres (see Fig. \ref{fig:mobility}(b))  \cite{morresi2020exploring,Taioli_2016} having a few millions carbon atoms with a number of Stone-Wales defects (reproduced in the inset of Fig. \ref{fig:mobility}(c)). The presence of bumps and asymmetries of the bands around the Fermi level, different from the linear dispersion of graphene, reflects the presence of penta-heptagonal defects related to the negative curvature \cite{morresi2020exploring}. TB can also be used to determine the bandstructure of twisted bilayer graphene, that shows interesting flat-band effects when the twist angle is small and the respective unit cell very large -- too large to be treatable by conventional methods such as DFT (Fig.~\ref{fig-twisted}).

\subsubsection{Limitations}
The TB approach relies on a basis set of atomic-like orbitals, and how efficient such approximation is depends on how many terms of the expansion are necessary to describe the property of interest. The main objection to the use of TB is that the electron wavefunctions within a periodic solid are expressed as a combination of atomic orbitals that are eigenstates of the Schrödinger equation for the constituent atoms, thus with a different potential and different boundary conditions with respect to the solid system. These local wavefunctions may  reasonably describe the states near the core of the atoms, while they are unsafe to represent a Bloch state in the insterstitial regions, where the electronic behavior is better described by a linear combination of plane waves. This is by the way the rationale of DFT approaches based on linear augmented plane wave (LAPW) basis sets with local-orbitals (such as in the ELK code suite \cite{elk}), whereby the solutions obtained within the Muffin tin spheres representing the core wavefunctions in the vicinity of the atomic region are matched to the plane-wave solutions in the interstitial region. In general, the TB approximation is not accurate to describe  high-energy valence states (not to say the scattering states), and more sophisticated methods are required, as we described in the following sections.

\subsection{Quantum transport: Landauer-B\"uttiker\label{LB}}

The LB formalism computes the conductivity of a sample by using a tripartite quantum junction model (see Fig. \ref{fig:mobility}d). In particular, the study of coherent mesoscopic transport consists of a system connected to a thermal reservoir by two reflectionless macroscopic contacts that define the temperature and the chemical potential of the incoming electrons (see Fig. \ref{fig:mobility}(d)). The transmission of electrons is assessed by solving directly the one-particle Schr{\"o}edinger equation in the junction region, which acts as a bottleneck for the electron current, while the left and right leads enter with separate Hamiltonians characterised by two distinct sets of eigenstates representing states filled according to different chemical potentials. \\
\indent  
Using this layout, one derives the so-called two-probe LB conductivity formula. This formalism can be used both within the  quasi-particle Fermi liquid framework and within correlated electrons schemes, when the Coulomb interaction between the flowing electrons breaks down the Landau picture. 

In the LB theory, the conduction through the device is represented in terms of scattering processes that the electrons undergo after being injected from the left lead into the device, and before entering the reservoir by crossing the right lead.
 Within the LB scheme the conductance reads:
\begin{equation}\label{buttiker}
I=\frac{2e}{h}\int 4 {\bf Tr}[\Gamma_L G_D^{r\dagger}\Gamma_R G^r_D][f(E-\mu_L)-f(E-\mu_R)] dE
\end{equation}
where $T(E)=4 {\bf Tr}[\Gamma_L G_D^{r\dagger}\Gamma_R G^r_D]$ is the lead-to-lead transmission probability of an electron of energy $E$, the integral is over all available energies, and the factor 2 counts the spin multiplicity. The intrinsic energy linewidths $\Gamma_{L/R}=\frac{i}{2}[\Sigma_{L/R}^r-\Sigma_{L/R}^{r\dagger}]$ of the left and right leads account for the finite lifetime of the electrons moving from the central region, where the conductance is described in terms of the retarded Green’s function (rGF) of the device ($G_D^r$), to the leads. 

\begin{figure}
   \includegraphics[width=8cm]{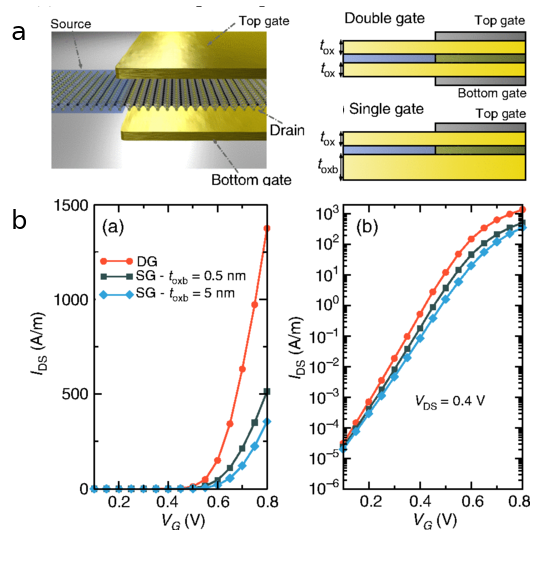}
  \caption{{\bf Multi-scale simulation of a MoS$_2$ transistor device}. The simulation of a `planar barristor' is based on DFT, Green's function formalism and Landauer-B\"uttiker formula. {\bf a}$|$ Transistor scheme in single gate (SG) and double gate (DG) configuration. {\bf b}$|$ Transfer characteristics in linear/semilogarithmic scale of the DG planar barristor, and of the SG planar barristors with different oxide thicknesses. $I_{DS}$, $V_G$ and $V_{DS}$ stand for the drain-source current, gate voltage and drain-source voltage, respectively. From REF.~\cite{PhysRevApplied.8.054047}, Copyright (2017) by The American Physical Society.
\label{fig-device}
}
\end{figure}

The main tasks of the LB approach consist thus in the accurate evaluation of the self-energies associated with the left and right electrodes, and of the Green's function (GF) of the central region (the device). These tasks can be accomplished via two different computational schemes: (i) GF evaluation recursive methods, similar to the techniques used in connection to TB Hamiltonians \cite{PhysRevB.28.4397,Sancho_1985,PhysRevB.28.6896}; (ii) {\it ab-initio} approaches~\cite{PhysRevB.59.11936,Rocha}. 
A schematic logic flow of the steps involved in the ab-initio simulation of the LB approach to transport is reported in Fig.~\ref{fig:mobility}(e).
Figure~\ref{fig-device} illustrates an {\it ab-initio} calculation of device voltage-current characteristics based on the LB approach.

\subsection{Quantum transport: Kubo approach} 

A second widely used quantum mechanical approach for coherent transport of electrons is based on the Kubo formalism \cite{Kubo_1966}. The Kubo approach is a broadly applicable technique, based on the linear response of a material to an externally applied electric field and on the fluctuation–dissipation theorem. In quantum transport \cite{PhysRevLett.100.036803,PhysRevLett.79.2518,Triozon}, this theorem relates the conductivity $\sigma(\omega)$,  ie. the dissipative out-of-equilibrium response at frequency $\omega$ (which can be derived from the current density  $\mathcal{J}(\omega)=\sigma(\omega)E(\omega)$, where $E(\omega)$ is the applied electric field), with the correlation function of the charge carrier velocities, which measures the fluctuations that the system undergoes by applying e.g. an external (weak) electric field.
In 2D materials, these fluctuations correspond to electronic transitions between states of the system at equilibrium induced by an oscillating field $E(\omega)=E_0 \cos(\omega t)$, which are connected to the total power absorbed per unit time and volume \be\mathcal{P}=\mathcal{J} \dot E=\sigma \langle E \dot E \rangle=\sigma E_0^2/2.\ee 
Using this theorem, at first-order perturbation theory in the electric field one obtains the Kubo conductivity  \cite{Kubo_1966}:
\begin{equation}\label{Kubo}
\small
\sigma(\omega)=\frac{\pi\hbar e^2}{\Omega}\sum_{m,n}|\langle m|v_{pl}|n\rangle|^2 \delta(E_m-E_n-\hbar\omega)[f(E_n)-f(E_m)]
 \end{equation}
%
where $\Omega$ is the sample volume,  $E_m,E_n$ are the energies of the levels $m,n$, $f(E_{n,m})$ is the Fermi–Dirac distribution, $e$ is the electric charge, and $v_{pl}$ is the projection of the velocity operator $\bf{v}$ on the axial direction within the 2D plane. 

The calculation of the Kubo conductivity is computationally very
expensive owing to the numbers of orbitals that must be included in the simulation.
However, efficient real-space implementations have been
 proposed in Refs. \cite{PhysRevLett.79.2518,morresi2020exploring,Roche}. 
The Kubo approach has been applied for example to simulate transport characteristics of several carbon-based materials \cite{ISHII2009283,doi:10.1021/nl0514386,PhysRevLett.100.036803,PhysRevLett.101.036808,PhysRevB.59.2284,PhysRevLett.87.246803,PhysRevLett.95.076803} (Fig.~\ref{fig:mobility}d,e). 
 
\subsection{Semiclassical approaches: Boltzmann Transport Equation}

Semiclassical approximations can be adopted when the electrons undergo incoherent or phase-breaking scattering events within the system and the connection of the leads to the electrodes is weak enough so that the charge inside the sample is well defined and quantified. 

Typically, this sequential regime is modelled by rate equations, such as the following BTE:
\begin{equation}\label{BTE}
\left[ \frac{\partial}{\partial t} +\bf{v}_k \cdot \nabla_{\bf{r}} +\bf{F} \cdot \nabla_{\bf{k}} \right] f_k({\bf r},t)=
\frac{\partial f_k(\bf{r},t)}{\partial t}|_{\rm coll}
\end{equation}
in which the quantum particle dynamics, under the external force-field $\bf{F}$, is described via a distribution function or particle density $f_k(\bf{r},t)$.
The term in the right hand side of Eq. \ref{BTE} accounts for particle-particle collisions that drive the system toward equilibrium. 

To made accessible the calculation of the conductivity, the solution of the BTE \cite{Roche} can be found within the Relaxation Time Approximation (RTA), which assumes that 
\be \frac{\partial f_k({\bf r},t)}{\partial t}|_{coll}=-\frac{g_k}{\tau_k},\ee 
where $g_k=f_k-f_k^{eq}$ represents the variation of the distribution function $f_k(\bf{r},t)$ from the equilibrium Fermi-Dirac distribution $f_k^{eq}({\bf r},t)$ at temperature $T$. The relaxation time $\tau_k$ is the time that the system takes to relax to $f_k^{eq}(\bf{r},t)$ after the external force-field is switched off.

The time-scale $\tau_k$ is typically assumed to be inversely proportional to the probability of scattering from the momentum state $|k'\rangle$ to $|k\rangle$, as given by Fermi’s golden rule. Finally, the calculation of the Boltzmann conductivity $\sigma$ within RTA is derived from the current density as:
\begin{equation}
\sigma=\frac{-e^2}{2\pi}\int kdk \left(  \frac{\partial f_k^{eq}(\bf{r},t)}{\partial \epsilon_k} \right ) \tau_k v_k^2
\end{equation}
where $\epsilon_k$ can be obtained via TB model simulations. 
Note that the BTE is a semiclassical equation, neglecting quantum interference between particles.

\subsection{Semiclassical approaches:  Monte-Carlo}

Carrier transport in 2D materials can be also modelled by using a MC statistical method, where electrons or holes are point-like classical particles moving along trajectories determined 
by elastic and inelastic scattering events \cite{taioli2010electron,DAPOR}, whose cross sections are assessed via quantum-mechanical approaches.
The MC approach may deal with several different possible scattering mechanisms: (i) elastic scattering with atomic nuclei, producing angular deviation of the electron/hole trajectories; (ii) inelastic e-e interaction leading to electron energy loss, secondary electron generation and angular deviation; (iii) electron-phonon interaction, typically introduced in a semi-empirical fashion at fixed energy loss; and (iv) trapping phenomena that end the trajectory (e.g. polaron quasi-particle excitation in insulating materials). 

The electron path is generally assumed to be described by a Poisson-like law, so the step length ($\Delta s$) between two subsequent collisions is given by:
\begin{equation}\label{mfp}
\Delta s = - l_\mathrm{tot} \ln(r), 
\end{equation}
where $l_\mathrm{tot}= \frac{1}{N \ \sigma} $ is the total mean free path, $N$ is the atomic number density, and $r$ is a uniformly distributed random number in the range $[0:1]$. \\
\indent Elastic scattering is typically described using either the Mott theory \cite{DAPOR,mott1929scattering,  1749-4699-2-1-015002,PhysRevB.79.085432} or first-principles approaches based on the numerical solution of the Dirac-Hartree-Fock (DHF) equation \cite{taioli2020relative}. Inelastic scattering is assessed by using the Ritchie dielectric theory \cite{Ritchie_PhysRev_1957}, in which the key issue is the calculation of the momentum dispersion of the energy loss function (ELF) \cite{taioli2020relative}.
To determine the fate of the electrons in the 2D sample, electron trajectories
are finally generated by comparing the probabilities of collisional events in different energy ranges with random numbers \cite{DAPOR}. 

Using the MC approach it is possible to simulate the anisotropic features of electron transport in layered materials.
An example is the calculation of the plasmonic spectrum in highly oriented pyrolytic graphite (HOPG)\cite{azzolini2018anisotropic} (see Fig. \ref{fig:mobility}(f)).

The electron-phonon interaction matrix elements, whose square values are directly related to the carrier-phonon scattering rate via Fermi’s golden rule, can be evaluated using a DFT/DFPT framework. Density functional perturbation theory \cite{RevModPhys.73.515} (DFPT) is used to obtain the phononic dispersion relations, based on the DFT electronic structure.
 This approach has been used e.g. to estimate the temperature-dependent intrinsic scattering rates as a function of electron/hole energy in layered semiconductors.
For example, such calculations have been used to 
clarify the reason for the lower carrier mobilities measured in transition metal dichalcogenides (TMDCs), compared to graphene, and the regimes where the mobility is defect-limited \cite{PhysRevB.100.115409} or phonon-limited \cite{PhysRevB.90.045422}.  However, the multivalley nature of the band structure requires the inclusion of spin-dependent functionals to account for the spin-orbit band splitting, which is strong in TMDs and can affect charge transport, particularly in the valence band. Additionally, many-body approaches beyond DFT should be used for an accurate assessment of the band energies, which can have a severe bandgap underestimation in DFT.

\subsection{Limitations to the applicability of transport approaches for different regimes}

The applicability limits of each of the different approaches to charge transport in 2D materials discussed previously are essentially ruled by the comparison of the characteristic size ($L$) of the system with the relevant transport length scales, which are related to both the characteristics of the system and the conditions under which it functions\cite{di_ventra_2008}. 

\indent In this respect, i) the Fermi wavelength $\lambda_F$ (e.g. $\approx 35-100$ nm in graphene \cite{torres2014introduction}) defines the de Broglie wavelength at which quantum states may interfere; ii) the elastic mean free path $\lambda_{el}$ determines the mean distance that a particle can travel before the direction of its momentum gets randomized by phase-preserving elastic scattering events; iii) the phase coherent length $\lambda_{\phi}$ (e.g. $\approx 0.5  \mu$m in graphene \cite{torres2014introduction}) describes the mean distance between two successive events modifying the phase of the quantum wavefunction; iv) and, finally, the localization length $\lambda_{loc}$ gives the average spatial extent of quantum states. \\
\indent The relative magnitude of these characteristic length scales identifies the different transport regimes, which correspond in turn to the different theoretical and computational models described in the previous sections. Of course, this identification can neither be sharp nor exhaustive, as several modes can coexist simultaneously making this problem extremely complex. The limits and the scope of each transport method and of the relevant formalism will be the subject of this section. \\

\indent For $\lambda_F \simeq L \ll \lambda_{\phi}, \lambda_{loc}, \lambda_{el}$, the system shows a spectrum of discrete levels. 
Semiclassical methods, e.g. based on the BTE, successfully describe this regime as one can neglect quantum interference effects or the many-body electron-electron interaction.\\	

\indent For $\lambda_F, \lambda_{el}, \lambda_{\phi} \ll L, \lambda_{loc}$, the system is characterised by a continuum spectrum. As phase coherent mean free path is small compared to system characteristic length, electrons lose their quantum mechanical phase due to a number of inelastic scattering events (electron-electron or electron-phonon scattering) that lead to energy relaxation and, eventually, to decoherence of the wavefunction. With dephasing that occurs at a scale much smaller than the length of the conductor, $L \gg \lambda_{\phi}$, the Ohm’s law is valid, and the resistance is linear with the sample length. In this respect, this transport regime can be dealt with semi-classical (or classical) approaches, such as the BTE (or MC).

Within the BTE semi-classical framework, the quantum mechanical treatment deals with the fact that electrons move at a length scale comparable to their wavelength, thus they sense the system periodicity and behave like Bloch waves in the periodic potential of the ion arrangement. Quantum description enters also in the collision integral of the BTE (see Eq. \ref{BTE}) and in the use of the Fermi-Dirac statistics. Of course, the collision integral is an approximate model of the inter-particle correlations. Thus, approaches based on BTE lack time-reversal symmetry in the long time limit as the degrees of freedom related to the correlations are neglected with a loss of information \cite{di_ventra_2008}. The classical character enters in this approach as the applied field and the electronic motion induced in response to it are treated classically, thus by neglecting the quantum interference effects of the electron-electron and electron-impurity scattering. Thus, the BTE approach works well to describe e.g. the electron/hole transport in weakly disordered conductors.\\

\indent When $\lambda_{\rm el} < \lambda_{\phi}, \lambda_{\rm inel}$, only a fraction of the scattering events destroys the phase.
In this transport regime, the picture is dominated by multiple scattering events that  randomize electron momentum and are responsible for electron energy loss,  
and electrons eventually achieve a steady transport regime by losing memory of their initial conditions.  Ohm’s law is applicable with resistance increasing with $L$ (Ohmic regime).
In this regime, classical charge transport MC \cite{DAPOR} can be used. 

In charge transport MC particles are treated as point-like objects and their motion is described in terms of trajectories. This means that this method is particularly valid for dealing with fast-moving particles, whose the De Broglie wavelength is small with respect to inter-particle distance. In particular, it was demonstrated that the use of the concept of trajectories is questionable below 15 eV \cite{LILJEQUIST201445}. Also short range order may lead to diffraction events, where the wave nature of particles emerge, limiting the use of the transport MC method to situations where the scattering is prevailingly incoherent. Furthermore, in classical MC calculations the effect of the electron-electron interaction on the trajectories is neglected, and each electron is treated irrespectively of the others. Finally, being a statistical approach, a number of trajectories must be accumulated in order to reduce the noise to signal ratio, which can limit the accuracy. A typical number of trajectories to achieve acceptable accuracy to compare with experimental measurements is in excess of $10^9$. 

 We point out that the detailed knowledge of the main interactions and energy loss mechanisms that electrons suffer while moving within the target can be a cumbersome task and, indeed, represents an intrinsic computational burden of transport MC modelling. 
 Generally, the calculation of the elastic cross section relies on the self-consistent solution of the Dirac-Hartree-Fock (DHF) equation in order to account for both direct and spin-flip terms upon scattering \cite{DAPOR}. The numerical solution of the DHF equation for extended systems with heavy atoms is computationally expensive \cite{doi:10.1002/adts.201870030,taioli2020relative}. In this respect, one can rely on semi-empirical calculations using the screened Rutherford cross section \cite{DAPOR}. \\
 On the other hand, the assessment of the electron-electron inelastic cross section can be carried out either by using experimental data, typically available only in the optical limit \cite{azzolini2017monte}, or via first-principles simulations, such as using time-dependent density functional theory (TDDFT) \cite{azzolini2017monte,taioli2020relative}. The latter approach may reckon the dielectric response of materials also at finite momentum transfer to take into account the materials dispersion relation. However, this task can be prohibitively expensive for systems with large unit cells. 
 
The inclusion of other degrees of freedom, such as phonons or polarons, further increases the complexity and negatively affects the computational scaling with system size, unless these interactions are treated semi-empirically \cite{DAPOR}. Moreover, the inclusion of the exchange-correlation interaction between the primary beam and the target in the interpretation of EELS experiments is problematic \cite{pedrielli2021electronic}. Indeed, the electron inelastic mean free path (IMFP) is assessed within the first Born approximation neglecting both relativistic effects and exchange interaction between the primary and the target electrons. In principle, the exchange interaction can be included by means of the Born-Ochkur exchange factor \cite{Fernandez_Varea_1993,deVera2021}. However, this correction only affects the calculated IMFP at low energies of the primary beam ($< 500$ eV) \cite{Fernandez_Varea_1993,deVera2021}.

 The burden of performing MC calculations can be substantially lifted off if one uses the continuous slowing down approximation \cite{DAPOR}, whereby electrons can continuously lose their kinetic energy in their way within the material. However, this picture works only if the statistical fluctuations of the several different energy loss mechanisms that the electrons undergo are not of paramount importance in the description of the observable of interest. For example, the backscattering coefficient is not crucially affected by this strategy to deal with the energy loss, while one must rely on the more accurate energy-struggling approach for simulating the secondary emission of electrons from the material, whereby each of the energy losses occurring along the trajectory must be taken rigorously into account at a significant computational cost \cite{DAPOR}. 
 
Classical and semiclassical descriptions fail to describe transport regimes where the phase of the electronic wavefunctions, which is delocalized all over the characteristic dimension of the system, is preserved over long distances owing to low scattering rate with magnetic impurities, phonons or electrons. This is the case of systems characterised by high periodicity, and by a low density of impurities or defects. In this regard, the wave nature of charges becomes pivotal and quantum interference effects start to play an important role. In this case, the scattering is coherent and, thus, full quantum mechanical approaches must be used to deal with this regime. In particular, at $\lambda_F \ll L \ll \lambda_{\rm el}, \lambda_{\phi}, \lambda_{\rm loc}$ the sample size is smaller than the elastic mean free path. The resistance is independent of $L$ and only depends on the contacts. The system is in the regime of ballistic transport. Furthermore, for $\lambda_F, \lambda_{el} \ll L \ll \lambda_{\phi}, \lambda_{\rm loc}$ electrons undergo momentum relaxing scattering events while sample lies in a coherent transport regime. Fully quantum-mechanical approaches, such as the Kubo linear response theory, the LB, or the non-equilibrium Green function formalism \cite{camsari2021nonequilibrium,datta_1995} must be adopted, the choice depending on the real experimental conditions. \\

In the LB theory, as described in section~\ref{LB}, the conductance from the left to the right lead is assessed from the quantum scattering theory (thus capturing the quantum effects on the electron transport) using a non-perturbative expansion of the transition operator. 
%
The LB formalism is best suited to deal with charge transport in mesoscopic phase-coherent devices with multi-terminal electrical contacts \cite{Marmolejo_Tejada_2018,doi:10.1080/14786437008238472,PhysRevB.31.6207,Sancho_1984} even though it can also handle bulk systems \cite{5390006,PhysRevB.40.8169,PhysRevB.64.165303} with the proper choice of self energies. In this regard, it has a direct connection with two-point transport measurements.
However, within the Landauer formalism, typically the external leads do not interact dynamically with the sample \cite{di_ventra_2008}. 
Unfortunately, this specific physical assumption may not be fulfilled in real experimental conditions as it means that the system is required to evolve into an incoherent set of independent scattering channels in thermal equilibrium rather than being a real open system interacting with its environment. 
Indeed, it requires the existence of well defined asymptotic conducting channels. 
Furthermore, in the LB approach the current is stationary at any time, which means that dissipation and thermalization of electrons takes place in other parts of the circuit. This assumption causes a loss of information related to the dynamical part of the current density, which is present in real devices. Also, the tripartition of the system in the junction and leads is somehow arbitrary and typically dictated by the computational tractability of the problem. In fact, a major limit of the Landauer formalism arises when one deals with arbitrary geometry.

To overcome these limitations, one can resort to the Kubo approach. This method works under the assumption that the system where the current flows is in global equilibrium, while small perturbations, which in the case of charge transport simulations are electric fields applied sufficiently slowly to the sample, allow one to keep only the linear terms in the electromagnetic vector potential expansion of the many-body Hamiltonian. 
In other words, the response of the system is expressed in terms of its equilibrium properties. This assumption, which is based on the fluctuation-dissipation theorem, limits the use of Kubo formalism to non-equilibrium states close to equilibrium. 
Furthermore, within the Kubo approach to charge transport the time evolution of the system is assumed to be faster than that induced by the external bath (``adiabatic approximation'') \cite{di_ventra_2008}. Thus, Kubo formalism is hardly applied to describe the long time or d.c. limit (frequency $\omega \rightarrow 0$), where the presence of the bath may drive the system towards equilibrium, thus invalidating one the most basic assumption of this approach.  The Kubo formula can be used to extract transport coefficients in the presence of many-body interactions \cite{PhysRevB.71.224423,Bohr_2006,PhysRev.127.5}. 

The Kubo and the LB approaches can be reconciled and are actually equivalent in some given conditions, most notably if the quantum transmission at the system/electrode interface is perfect.
However, the Kubo approach is better suited to investigate the transport properties of disordered materials, such as those with vacancies and magnetic impurities, characterized by localization phenomena in the low-temperature limit  \cite{PhysRevLett.98.076602}. 
In this respect, we point out that while in the delocalized regime ($L \ll \lambda_{\rm loc}$) the individual electronic states spread from one contact to the other and coherent transport of electrons is achieved, for $\lambda_F, \lambda_{\rm el}, \lambda_{\rm loc} \ll L \ll \lambda_{\phi}$ constructive interference can lead to the localization of electrons in closed scattering loops, resulting in a drop of the conductivity. This regime, where coherent transport within a single state is exponentially suppressed, typically induced by the presence of disorder (disorder-induced Anderson localization regime \cite{PhysRev.109.1492}), results in strong localization of the electric charge. In this regime conductance can only originate from thermally activated hopping between electronic levels.

In the LB approach the calculation of the transmission function relies on the assessment of the Green’s functions of the full system, whose calculation needs cubic-scaling matrix inversion techniques that make the LB formalism computationally prohibitive for large and disordered 2D and 3D systems.
At variance, a few tailored effective linear scaling quantum transport numerical methods have been developed specifically in the context of the Kubo formalism \cite{FAN20211} and attain their full capabilities within this framework. This computational scaling advantage explains the reason behind 
the application of Kubo for studying transport in disordered materials, including 3D metals and 2D Dirac semimetals.

Finally, a fundamental limitation is represented by the computational cost for a given accuracy.
In low-dimensional materials, where localization effects are more pronounced, an accurate treatment of quantum coherence on electronic transport is crucial \cite{PhysRevB.89.075420}. Some of the most interesting effects in conductivity are caused by the electron-electron interaction and cannot be described by perturbative methods. One prominent example is the Kondo effect that is typically observed in metals \cite{10.1143/PTP.32.37} but has also been found in CNTs \cite{cite-key}, another one being the physics of the Luttinger liquid \cite{doi:10.1063/1.1704046}. To lower the computational load, Coulomb interaction between electrons can be captured occasionally by a mean field approximation, and decoherence can to some degree be mimicked by non-Hermitean terms in the Hamiltonian. However, for an accurate treatment of the many-body interaction one must rely on more sophisticated many-body approaches. 
We also notice that although both LB and Kubo are fully quantum mechanical formalisms, they reproduce the semiclassical BTE to leading order in $\lambda_{F}/\lambda_{\rm el}$ for completely incoherent systems, where quantum interference between individual states can be neglected.

\section{Mechanical properties} \label{MECH}

While the optical, electrical, and transport properties of 2D materials are extensively investigated for their potential application in microelectronics, many of the current commercial applications of 2D materials rely on their mechanical strength. In addition, 2D materials are also expected to play a pivotal role in flexible electronics \cite{C9NR03611C}, and therefore their mechanical properties are of major interest. 
Theoretical approaches for studying mechanical properties can be  categorised into three classes: (1) continuum mechanics (CM), e.g. based on finite elements (FEM) methods, and peridynamics \cite{signetti20172d}; (2) atomistic simulation based on classical molecular dynamics (MD); (3) ab-initio simulations based on DFT or TB.

DFT yields accurate elastic constants, especially those corresponding to in-plane deformations, as covalent bond energies are quantitatively more accurate than vdW energies.
DFT is typically applied to defect-free materials with a limited number of atoms in the unit cell, lower than 100. 
The elastic constants can be calculated from the total energy at 0~K, for selected deformations.
For example, Fig. \ref{fig:mobility}(g) illustrates the simulation, using DFT, of the stress-strain characteristics of several 2D carbon allotropes with density lower than graphene \cite{morresi2020structural}.
 
Nevertheless, the mechanical properties of real materials are often dominated by defects, such as dislocations, and vary considerably from those of the perfect crystals. Thus, models at larger scale are needed. One alternative is represented by semi-empirical TB simulations \cite{doi:10.1063/1.5143190}, which were for example used to simulate armors based on carbon or hybrid layers \cite{signetti20172d}. In that study, a size-scale transition from the nano- to the microscale in the impact behaviour was found by adopting a multiscale approach based on FEM and CM models. 

For larger computational systems, in a space-time scale of $\approx 1 \mu$m  and $\approx$~1 ns, classical MD simulations are mainly used. In classical MD simulation, electrons are not explicitly included in the equation of motion, and the interatomic potentials between atoms in the systems are pre-calculated and tabulated for several bond orders and lenghts, and typically modelled by two-body or reactive semi-empirical force-fields \cite{Arduin}, such as condensed-phase optimized molecular potentials (COMPASS) \cite{doi:10.1021/jp980939v}, reactive empirical bond order (REBO) potentials \cite{PhysRevLett.63.1022}, or adaptive intermolecular reactive empirical bond order (AIREBO) potentials \cite{doi:10.1063/1.481208}. These proved effective in the analysis of the mechanical behaviors of several carbon-based 2D materials, e.g. strained graphene \cite{doi:10.1021/nl901448z,SAKHAEEPOUR200991,SAKHAEEPOUR2009266,doi:10.1063/1.4747719,Zheng_2011,PhysRevB.81.235421}, and carbon-based foams \cite{pedrielli2018mechanical,pedrielli2017designing}. 
However, parameterised potentials are not readily available for all element combinations and types of bonding, and substantial effort may be required to set up and test a classical simulation for a new material. Today, machine learning, trained on DFT data, is also used to generate these empirical potentials \cite{van_der_Giessen_2020}.

\subsection{Limitations}
Mechanical simulations of phenomena that involve processes at multiple length or time scales are still a challenge. The error introduced by combining different multiphysics approaches (eg. DFT and classical potentials) is difficult to minimize and to estimate. For this reason, material modeling problems at mesoscopic or microscopic length scales (eg. graphene fiber aggregation) and that at the same time involve large deformations, phase transitions, hysteresis, or the combination of mechanical deformation with chemical and electronic phenomena are also amongst the most technically challenging.

\section{Ferroelectric properties\label{ferro}}
Ferroelectric materials have a spontaneous electric polarisation that can be reversibly switched by application of an external electric field. First principles calculations based on density functional theory have been used to predict the existence of ferroelectric phases for 2D materials\cite{hanakata2016polarization,c2020electronic,qi2021review}, where phenomenological theory has limited application due to its continuous description.  Contrary to the previous belief that ferroelectricity vanishes for thin films \cite{merz1956switching}, ferroelectric monolayer materials have been realised experimentally \cite{higashitarumizu2020purely,chang2016discovery}, even in cases where the bulk is not ferroelectric \cite{lee2020scale}.
This is the case of HfO$_2$, which exhibits an out-of-plane polarisation at room temperature in thin films \cite{lee2020scale}. Due to the unusual spacing between dipoles, the domain wall formation energy is nearly vanishing and the cohercive electric field is very high. Other ferroelectrics such as group-IV monochalcogenides with SnS structure are centrosymmetric in bulk or for systems with even number of layers, but polar for an odd number of layers \cite{hanakata2016polarization,c2020electronic,chang2020microscopic}. Exceeding the expectations from theoretical predictions, 
ferroelectricity was experimentally found both for SnS thin films with odd and even number of layers \cite{higashitarumizu2020purely}.

Essential steps in the prediction of ferroelectric phases are 
(1) determining the existence of a crystal phase with spontaneous polarisation, and
(2) quantifying the coercive electric field, or alternatively the energy, necessary to switch the direction of polarisation. 

The appearance of a spontaneous polarization during a ferroelectric phase transition is a consequence of a structural change from a non-polar to a polar symmetry (usually upon cooling).
Using DFT calculations, it is possible to verify that the symmetry of the low-energy structure is polar, and to search a centrosymmetric structure into which the polar structure can be transformed by a continuous change of an order parameter. 
As for the choice of the order parameter, the polarization is usually adopted \cite{tagantsev2008landau} in the Landau description \cite{chandra2007landau}. Another option in some cases \cite{tagantsev2008landau}  is to use the soft phonon-mode displacement. The soft phonon can be determined by calculating the phonons of the centrosymmetric structure.
The energy barrier associated with the polarisation switching can be obtained using the nudged elastic band method\cite{henkelman2000climbing}, and the respective coercive field by relaxation under an application of an electric field. 

For ferroelectrics with polarization change perpendicular to the layer, such as few-layer HfO$_2$ \cite{lee2020scale} or monolayer PbS \cite{hanakata2018strain}, the electric field can be trivially included in the DFT calculation by chosing an external potential
that is commensurate with the periodic boundary conditions (if there are any), such as a sawtooth potential. 
In-plane electric fields can be introduced in supercell DFT calculations by employing a perturbative approach \cite{PhysRevLett.89.117602,nunes2001berry}
in combination with the modern theory of polarization \cite{vanderbiltPRB1993}.

The polarization is given by
\begin{eqnarray} {\bf P}=\frac{1}{V}\sum_iq_i^{\rm ion}{\bf R}_r-\\ \frac{2ie}{(2\pi)^3}\sum_n^{\rm occ}
\int_{\rm BZ}d^3{ k}  \langle\Psi_{n{\bf k}}|\frac{\partial \Psi_{n{\bf k}}}{\partial {\bf k}}\rangle,  \end{eqnarray}
where $q_i$ is the ionic charge including the core electrons, $R_i$ is the position of ions, $V$ is the unit cell volume, $e$ is the elementary charge, and $\Psi_{n{\bf k}}$ are the periodic electronic wavefunctions of the occupied states (the Kohn-Sham states).
The first term is the ionic contribution.  Note that this is not a well defined quantity for an infinite crystal as it depends on how the unit cell is chosen, and therefore the {\it change} of polarisation needs to be calculated \cite{tagantsev1991electric,PhysRevLett.69.389}. The second term is the electronic contribution  which can be calculated from the Berry phase \cite{vanderbiltPRB1993}. A modified Berry phase calculation is needed to evaluate the polarization of ferroelectric metals \cite{filippetti2016prediction}.

\subsection{Limitations}

Ferroelectricity depends mostly on ground-state properties that are well described within DFT. However, some structures may suffer of particular shortcomings, such as group-IV chalcogenides, for which the calculated lattice parameters and relative energies vary greatly with the functional chosen \cite{barraza2018tuning}.
Finally, the determination of phase transition temperatures remains a challenge for ferroelectrics and other phase transitions alike.

\section{Magnetic properties\label{magnetic}}

Despite the previous belief, based on an overgeneralisation of the  M-W theorem  \cite{mermin1966absence,halperin2019hohenberg}, that ferromagnetic or antiferromagnetic order in 2D materials would be impossible above 0~K, 2D ferromagnetic materials have also been predicted theoretically and synthetised experimentally \cite{gong2019two}.

Examples of 2D magnetic vdW heterostructure materials are CrI$_{3}$, CrCl$_{3}$, Fe$_{3}$G$_{2}$Te$_{2}$, VS$_{2}$, VI$_3$, Cr$_2$Ge$_2$Te$_6$ and FePS$_3$ \cite{huang2017layer,jin2018raman,wu_physical_2019_CrI3,CrI3_Su_rez_Morell_2019,Kashin_2020_CrI3,VI3,kutepov2021electronic,kvashnin2020dynamical,gong2017discovery,lee2016ising}
and many others have been predicted theoretically \cite{cortie-advanced-review,gong2019two}.
 Moreover, many of these systems show different magnetic phases depending on the number of the layers, generating widely interest for possible technological applications.
Prediction of ferromagnetic and anti-ferromagnetic phases often relies on spin-polarized DFT (see section~\ref{DFT}),
which can be used to calculate the spin magnetization and the
magnetic  energy, the difference in energy between the non-spin-polarized and ferromagnetic  ground  states. 
However, it must be noted that many density functionals favor magnetic states \cite{fu2019density}.
DFT has also been used to predict itinerant magnetism in doped 2D semiconductors \cite{SnO,das2018electronic}.

Non-collinear magnetic states however, require employing a formulation of DFT for vector spins \cite{kubler1988density}. Spin-orbit interactions,  essential to account for magnetic anisotropy, can be obtained using a relativistic DFT formalism \cite{fernandez2006site,fernandez2006siteerratum,cuadrado2018implementation}. 

Prediction of phase transitions involving magnetic ordering is often described using an Ising model, if spins are oriented along the out-of-plane direction, or using a Heisemberg model for arbitrary spin orientations. CrI$_3$ can be approximately described using a 2D Ising model\cite{huang2017layer}.
However, to describe the magnetic anisotropy it is necessary to consider spin canting. This can be accomplished by using effective spin models based on the Heisenberg model with additional terms \cite{lado2017origin} including the
Dzyaloshinski-Moriya term or antisymmetric  exchange $\vec{D}_{ij}\cdot(\vec{S}_i\times\vec{S}_j)$ \cite{moriya1960anisotropic}.
The Dzyaloshinski-Moriya  interaction favours spin-canting and often plays an important role in stabilizing skyrmions.

Magnon bands can be simulated, for example, by using many magnetic configurations, or using a classical Heisenberg model with  magnetic interaction parameters obtained from DFT \cite{LIECHTENSTEIN198765,HE2021107938,cuadrado2018implementation}.
In a recent article \cite{Katsnelson_e_electron_2021}, a TD-DFT study of bulk CrI$_3$ has shown a gap opening of $1.8$ meV in the absence of spin-orbit coupling, due to electron correlation effects. Also ab initio BSE with spin-orbit coupling (sec. \ref{BSE}) can be performed to calculate the absorption spectrum and the magnon dispersion in 2D CrI$_3$ \cite{olsen2021unified}. In these results, spin orbit coupling induces a gap between the acoustic and optical magnon branch of 0.3 meV, indicating possible DM interactions.

\subsection{Limitations}

Magnetism is one of the manifestations of electron-electron interaction, which is one of the hardest terms to estimate within DFT (see section~\ref{DFT}).
The emergence of magnetism in correlated electron phases can be treated for example using the DMFT formalism (section~\ref{DMFT_section}).

Nevertheless, band theory is still often used to predict the conditions under which magnetic states are expected to appear, due to the lack of an alternative formalism departing from the non-interacting picture but at the same time conceptually simple.
Take, for example, the emergence of magnetism on twisted graphene systems at fractional filling\cite{sharpe2019emergent,polshyn2020electrical}: the concept that when graphene is stacked at a small twist angle, the resulting bands near the  Fermi level become flat is essential to understanding why electron-electron interactions are strongly enhanced, but insufficient to predict the resulting interacting state.

Predicting the thermodynamic stability of magnetic states, competing highly correlated phases and superconductivity in 2D materials remains an open challenge from the point of view of both fundamental theory and numerical modelling. Shape and magnetocrystalline anisotropies, which play an important role in breaking the validity conditions of the Mermin–Wagner theorem and in stabilizing magnetic order, are of the scale of $\mu$eV \cite{cortie-advanced-review}. This small energy scale, together with the large associated length scale, make it challenging to compute magnetism in 2D materials.

\section{Outlook}

The revolutionary advances that 2D materials have known over the last ten years have been greatly supported by computational modelling. Conversely, the development of computational methods is often driven by the need to explain new experimental results.
A major motivation that drives experimental and theoretical efforts is to find new materials and devices able to achieve superior information density and higher energy efficiency with respect to conventional electronics, or to support flexible electronics. 
In parallel however, 2D materials investigations have been driven by the possibility of studying new physical phenomena.
This in turn brings new questions to be answered by theoretical and computational means.

First-principles approaches able to treat electronic excitations, based on TD-BSE and  TD-DFT, demonstrated their potential to interpret experimental outcomes. Nevertheless, {\it ab-initio} TD-BSE is computationally expensive and TD-DFT suffers from a kernel that does not take into account accurately the excitonic effects. In this regard, the computational development is focusing on numerical approaches that are able to accelerate the calculations.

Further methodological advances are also desirable in the computational treatment of the excitonic phenomena that have no counterpart in 3D. This is mainly due to the lack of dielectric screening outside the plane which stabilises excitons, trions, biexctions, etc.
Additionally, in vdW structures, the possibility of forming quasi-particles localised in different layers, such as inter-layer excitons, opens up intriguing possibilities.
For example, in heterostructures with layers stacked at a finite twist-angle, leading to moir\'e superlattices, it is possible to choose both the momentum displacement between the electron and hole bands and the confining potential. The computational description of the resulting moir\'e excitons, including new selection rules \cite{wu2018theory,yu2015anomalous} and trapped/delocalised phases \cite{brem2020tunable}, is an extremely active area of research.

The {\it ab-initio} description of exciton-phonon polaron QPs, characterised by Coulomb correlations renormalized by lattice dynamics via polaronic effects, is still 
in its infancy.
In particular, to achieve a complete theoretical treatment to predict the full excitonic dispersion in 2D materials is a complex challenge, as one would need to include in one unified framework several microscopic ingredients, such as spin–orbit coupling, exchange interactions, many-body correlations, and polaronic effects.
Non-equilibrium exciton dynamics and its effects in the optical and electronic properties of 2D materials, such as the carrier-density dependent optical spectra, are very challenging for first principles calculations but can be tackled using density matrix theory \cite{steinhoff2014influence,Nature_biexciton,selig2018dark}.

Computational modelling is still hampered by the widespread use of basis sets adopting 3D periodic boundary conditions, which are not naturally adapted to 2D systems. This approach leads to an unnecessary computational expenditure and, specially in GW, to reduced accuracy.
Furthermore, pristine 2D materials are unstable against the creation of flexural phonons and magnons (for infinite ferromagnetic systems), as a result of the  Mermin-Wagner-Hohenberg theorem. It is not surprising that these collective excitations are still challenging to model in 2D, owing to the difficulty in accounting for substrate dampening effects, or the numerical instability resulting from the lack of dampening in free layers.

For example, the description of effects induced by correlation in confined systems remains a challenge, owing to the difficulty in describing both the stronger electron-electron interaction and the decreased screening of external electromagnetic fields (compared to 3D).
Twisted (moir\'e) multilayers, as well as other 2D systems, such as FeSe monolayers, have recently revealed the emergence of many interesting electronic phases, including orbital magnetism, superconductivity, or quantised anomalous Hall effect
\cite{cao2018correlated,cao2018unconventional,yankowitz2019tuning,lu2019superconductors,serlin2020intrinsic,kreisel2020remarkable,shi2020tunable}.
Despite recent developments, an adequate computational modelling of highly correlated electronic phases in 2D materials is still lacking \cite{castroneto-RevModPhys.84.1067,Tang570}, particularly in presence of external magnetic fields, such as those routinely applied to probe these phases in experiments.

Ferromagnetic systems are also an emerging area within 2D materials research. 
Although there is at present no doubt that 2D ferromagnetic materials are stable, there is still lack of information on the role of boundary conditions, substrate, etc. in stabilising magnetism. Similar remarks apply to 2D ferroelectrics: for example, it is not completely understood the critical thickness dependence of ferroelectric polarization, as well as the relevant experimental conditions. Since ferromagnetism and ferroelectricity are very dependent on defects and boundary conditions, we believe that computational studies will be paramount to unravel these questions.
Other important emerging areas that we must mention are related to the modelling of magnetic topological excitations, such as skyrmions. 

We point out also the need for developing computational methods to calculate macroscopic observables that are scalable and computationally effective in terms of memory requirements and CPU usage, notably for modelling realistic systems.
There is still ample room for development of first-principles based multiscale approaches to calculate macroscopic observables \cite{van2020roadmap} and to link the properties of 2D materials to the measurable electrical response of real devices, such as carrier mobilities, thermal conductivity, ionic conductivity, spin lifetime, etc.
While computer simulations and experimental measurements have quantified the intrinsic carrier mobility of quite a few 2D  materials, its assessment in presence of substrates and external fields, at real-scale geometries, is still challenging for computer simulations.  Furthermore, as the size of devices is decreasing to the limit where quantum effects play a crucial role,  the computational modelling of transport properties of 2D materials, which are at the heart of the modern quantum revolution, must treat accurately the physics of coherent transport also incorporating electron-electron interaction into the simulation.

Additionally, many systems of technological interest are not crystalline but amorphous (such as graphene oxide), or are molecular 2D layers (such as 2D water). 

Finally, emerging areas of computational research in 2D materials modelling aim to machine learning models and inverse design \cite{mounet_two-dimensional_2018,thygesen_ML2d,zhou_2dmatpedia_2019,zunger2018inverse,ren2020inverse}. Machine learning is being used e.g. to revolutionise the fitting of classical inter-atomic potentials or other approximate models by replacing the considerable human effort with an artificial neural network \cite{mortazavi2020machine,vasudevan2020investigating}. This may change classical modelling approaches and make them promptly available in the area of application.

\bibliographystyle{iopart-num}
\bibliography{refs}

\end{document}